\documentclass[10pt,onecolumn]{IEEEtran}

\usepackage{flushend}
\usepackage[cmex10]{amsmath}
\usepackage{amsfonts}

\usepackage{algorithmic}

\usepackage{array}
\usepackage{epsfig}
\usepackage{bbm}
\usepackage{mdwmath}
\usepackage{mdwtab}
\usepackage{eqparbox}
\usepackage[font=footnotesize]{subfig}
\usepackage{amsfonts}
\usepackage{tikz}

\newtheorem{theorem}{Theorem}

\newtheorem{lemma}[theorem]{Lemma}
\newtheorem{corollary}[theorem]{Corollary}

\newtheorem{definition}{Definition}

\newtheorem{example}{Example}
 
\newcommand{\field}[1]{\mathbb{#1}}
\newcommand{\R}{\field{R}} 
\renewcommand{\Re}{\R} 
\newcommand{\E}{\mathbb{E}} 

\newcommand{\Fscr}{{\cal F}}
\newcommand{\Gscr}{{\cal G}}

\newcommand{\Lscr}{{\cal L}}

\newcommand{\Pscr}{{\cal P}}

\newcommand{\Xscr}{{\cal X}}

\newcommand\argmin{\mathop{\mbox{{\rm argmin}}}\limits}

\newcommand{\bea}{\begin{eqnarray}}
\newcommand{\eea}{\end{eqnarray}}
\newcommand{\beas}{\begin{eqnarray*}}
\newcommand{\eeas}{\end{eqnarray*}}
\newcommand{\twopartdef}[4]
{
	\left\{
		\begin{array}{ll}
			#1 & \mbox{if } #2 \\
			#3 & \mbox{if } #4
		\end{array}
	\right.
}

\begin{document}

\sloppy

\title{Relations between Information and Estimation in Discrete-Time L\'evy Channels}

 \author{
   \IEEEauthorblockN{Jiantao Jiao, Kartik Venkat and Tsachy Weissman\\}
   \IEEEauthorblockA{Department of Electrical Engineering\\
     Stanford University\\
     Email: \{jiantao,kvenkat,tsachy\}@stanford.edu} 
}
\maketitle

\maketitle

\begin{abstract}
Fundamental relations between information and estimation have been established in the literature for the discrete-time Gaussian and Poisson channels. In this work, we demonstrate that such relations hold for a much larger class of observation models. We introduce the natural family of discrete-time L\'evy channels where the distribution of the output conditioned on the input is infinitely divisible. For L\'evy channels, we establish new representations relating the mutual information between the channel input and output to an optimal expected estimation loss, thereby unifying and considerably extending results from the Gaussian and Poisson settings. We demonstrate the richness of our results by working out two examples of L\'evy channels, namely the gamma channel and the negative binomial channel, with corresponding relations between information and estimation. Extensions to the setting of mismatched estimation are also presented.  
\end{abstract}


\section{Introduction} \label{sec: intro}
Deep and elegant relations between fundamental measures of information and fundamental measures of estimation have been discovered for several interesting probabilistic models. Over time, such relations have been subject to interest in communities ranging from information theory to probability and statistical decision theory. For a recent comprehensive treatment of this topic and its implications, we refer to \cite{Guo--Shamai--Verdu2013}. While both discrete-time and continuous-time observation models have been extensively discussed in the literature, and intriguing interconnections between both regimes drawn, in this work we focus exclusively on the discrete-time case.  


Our story can be traced back to the early work by Stam \cite{Stam1959} in 1959, where ``de Bruijn's identity'' relating the differential entropy of a Gaussian noise corrupted random variable to its Fisher information was presented. However, a more concrete starting point is the recent work by Guo, Shamai and Verd\'u \cite{Guo--Shamai--Verdu2005} in 2005. In \cite{Guo--Shamai--Verdu2005}, the authors proposed the I-MMSE formula, which presents the derivative with respect to the signal-to-noise ratio (SNR) of the mutual information between the input and output of a Gaussian channel, as half the minimum mean squared error in estimating the channel input based on the output. Formally, if $X$, a random variable with finite variance\footnote{This condition can be weakened to that the mutual information $I(X;Y_\gamma)<\infty$ for some $\gamma>0$, as shown in~\cite[Thm. 6]{wu2012functional}.}, denotes the channel input, and $Y_\gamma = \gamma \, X + W_\gamma$ indicates the channel output at SNR level $\gamma > 0$, where $W_\gamma \sim \mathcal{N}(0,\gamma)$ is an independent Gaussian random variable, then the I-MMSE relationship can be stated as,
\begin{align} \label{eq: i-mmse}
\frac{\partial}{\partial \gamma} I(X; Y_\gamma) = \E[ \ell_{\Gscr} (X, \E[X|Y_\gamma]) ],
\end{align}
where the Gaussian loss function $\ell_{\Gscr}: \Re \times \Re \to [0, \infty)$ is defined as,
\begin{align} \label{eq: gaussian loss}
\ell_{\Gscr}(x, \hat{x}) \doteq \frac{1}{2} \, (x - \hat{x})^2.
\end{align}

In other words, for any choice of input distribution, the derivative  of the mutual information is equal to half the minimum mean squared error in estimation. It turns out that such a relationship between mutual information and optimal estimation loss is not unique to the Gaussian channel. Similar relations were found for the discrete-time and continuous-time Poisson Channel in \cite{Guo--Shamai--Verdu2008} and \cite{Atar--Weissman2012}. Remarkably, the exact same relationship holds in the Poisson context as well, when the squared error loss is replaced by a natural loss function for the Poisson channel. 

Indeed, consider a non-negative random variable $X$, satisfying $\E[X \ln X] < \infty$, and conditioned on $X$, $Y_\gamma \sim \mathsf{Poi}(\gamma \, X)$, now denote the Poisson channel input and output at SNR level $\gamma$, respectively. Invoking results from \cite{Guo--Shamai--Verdu2008} and \cite{Atar--Weissman2012}, we can express the relationship corresponding to (\ref{eq: i-mmse}) for the Poisson channel as, 
\begin{align} \label{eq: poisson i-mmle}
\frac{\partial}{\partial \gamma} I(X; Y_\gamma) =  \E [ \ell_{\Pscr}(X, \E[X|Y_\gamma]) ], 
\end{align}
where the Poisson loss\footnote{It turns out that conditional expectation minimizes the expected loss under $\ell_\Pscr$. For additional properties and discussion, the reader is referred to \cite{Atar--Weissman2012}.} function $\ell_{\Pscr}: [0, \infty) \times [0, \infty) \to [0, \infty]$ is defined as,
\begin{align} \label{eq: poisson loss}
\ell_{\Pscr}(x, \hat{x}) \doteq x \, \ln\left( \frac{x}{\hat{x}} \right) - x + \hat{x}.
\end{align}
The similarity between (\ref{eq: i-mmse}) and (\ref{eq: poisson i-mmle}) is quite striking. Indeed, the kinship between these two channel models does not end here. In \cite{Verdu2010}, Verd\'{u} extended the I-MMSE result to incorporate mismatch at the decoder. In this setting, the underlying clean signal $X$ is distributed according to $P$, while the decoder believes the true law to be $Q$. For the discrete-time Gaussian channel model with SNR level $\gamma$, which could be infinity, \cite{Verdu2010} presents the following relationship between the
relative entropy of the true and mismatched output laws, and the difference
between the mismatched and matched estimation losses: 
\begin{align}
D( P_{Y_\gamma} || Q_{Y_\gamma} ) = \int_0^\gamma \E_{P}[ \ell_{\Gscr} (X, \E_{Q}[X|Y_\alpha]) - \ell_{\Gscr} (X, \E_{P}[X|Y_\alpha]) ] \,d\alpha. \label{eq: gaussian discrete-time mismatch}
\end{align}
An essentially identical result was established by Atar and Weissman in \cite{Atar--Weissman2012} for the Poisson channel: 
\begin{align}
D( P_{Y_\gamma} || Q_{Y_\gamma} ) = \int_0^\gamma \E_{P}[ \ell_{\Pscr}(X, \E_{Q}[X|Y_\alpha]) - \ell_{\Pscr}(X, \E_{P}[X|Y_\alpha]) ] \, d\alpha, \label{eq: poisson discrete-time mismatch}
\end{align}
where, as in (\ref{eq: poisson i-mmle}), the overloaded symbol $Y_\gamma$ now denotes the output of the Poisson channel with input $X$ at SNR level $\gamma$. The first terms in the right hand sides of the integrands in \eqref{eq: gaussian discrete-time mismatch} and (\ref{eq: poisson discrete-time mismatch}) denote the average loss incurred when the decoder employs the estimator optimized for law $Q$. The right hand sides therefore indicate the cost incurred due to mismatch in estimation, integrated over a range of SNR values.   

Thus, we observe the connection between the Gaussian and Poisson observation models, wherein a direct relationship between mutual information, relative entropy and average estimation loss holds verbatim in both models, under the appropriate loss function. Further, the I-MMLE (Mutual Information-Minimum Mean Loss in Estimation) formulae stated in (\ref{eq: i-mmse}), (\ref{eq: poisson i-mmle}), and their mismatched D-MLE (Relative Entropy-Mean Loss in Estimation) counterparts in (\ref{eq: gaussian discrete-time mismatch}), (\ref{eq: poisson discrete-time mismatch}), hold for any choice of input distributions, as long as they satisfy benign regularity conditions. 

In this work, we aim to understand this special connection, and present a clear, unified picture assimilating both classical as well as unknown results in the world of information and estimation for a wide class of discrete-time observation models. Our main contributions here are fivefold:   
\begin{enumerate}
\item The introduction (to our knowledge, for the first time in the literature) of discrete-time L\'evy channels, which are a sub-family of the well-known generalized linear models \cite{Maccullagh--Nelder1989} in statistics. L\'evy channels satisfy the property that conditioned on the inputs, the outputs are random variables with infinitely divisible distributions. Additionally, they have a natural SNR parameter, which captures the channel quality. 

\item For discrete-time L\'evy channels, we present a simple relationship between mutual information and an optimal estimation loss. We also present the generalization of this result to incorporate mismatch at the decoder. Additionally, we provide new formulae for expressing the entropy and relative entropy in terms of estimation risks.  
 
\item We recover results for both the Gaussian and Poisson settings, for matched and mismatched estimation scenarios, as special cases of our general result. To our knowledge, this is the first unified presentation of information and estimation relationships for these two canonical discrete-time channels.

\item We present two natural channels, namely the gamma channel and the negative binomial channel, both of which are instances of L\'evy channels. For these channels, we use our general result to explicitly derive the information and estimation relationship.

\item We investigate the loss function that emerges in the characterization of mutual information in L\'evy channels. In particular, we show that when the inputs to the channel are deterministic values, the loss function reduces to a single Bregman divergence that is generated by the Fenchel--Legendre transform of the channel's cumulant generating function.     
\end{enumerate}

The remainder of this paper is organized as follows. In Section \ref{sec: levy-type channels}, we introduce discrete-time L\'evy channels as a natural discrete-time observation model, and discuss some properties underlying this family. In Section \ref{sec: main results}, we present our main theorems on relations between information and estimation for L\'evy channels. In Section \ref{sec: recover G and P}, we recover, as corollaries of our main result, the fundamental relationships already known for the Gaussian and Poisson channels. In Section \ref{sec: gamma}, we introduce and study two special L\'evy channels, namely the  gamma channel and the negative binomial channel, from an information and optimal estimation viewpoint. In Section \ref{sec: bregman} we discuss the natural loss function associated with L\'evy channels. We present the proofs of our results in Section \ref{sec: proofs} and conclude in Section \ref{sec: conclusions}.

\section{Discrete-time L\'evy channels} \label{sec: levy-type channels}
Note that $\ln(\cdot)$ denotes the natural logarithm, and $\sigma\{X_\alpha,\alpha\in \mathcal{A}\}$ denotes the smallest $\sigma$-algebra with respect to which the random variables $X_\alpha, \alpha \in \mathcal{A}$ are measurable.

\subsection{L\'evy processes and infinitely divisible distributions}

A general one-dimensional L\'evy process is defined as follows.
\begin{definition}[L\'evy process]
A process $Y = \{Y_t: t\geq 0\}$ defined on a probability space $(\Omega, \mathcal{F}, \mathbb{P})$ is said to be a L\'evy process if it possesses the following properties:
\begin{enumerate}
\item The paths of $Y$ are $\mathbb{P}$-almost surely right continuous with left limits.
\item $\mathbb{P}(Y_0 = 0) = 1$. 
\item For $0\leq s\leq t$, $Y_t - Y_s$ is equal in distribution to $Y_{t-s}$. 
\item For $0\leq s\leq t, Y_t - Y_s$ is independent of $\{Y_u: u\leq s\}$. 
\end{enumerate}
\end{definition}

Important examples of L\'evy processes include include Brownian motion and Poisson processes. We refer the reader to Sato \cite{Sato1999} for a comprehensive treatment of L\'evy processes. 

The infinitely divisible distribution is defined as follows:
\begin{definition}[Infinitely divisible distributions]
We say that a real-valued random variable $T$ has an infinitely divisible distribution if for each $n \in \mathbb{N}, n\geq 1$, there exists a sequence of i.i.d. random variables $T_{1,n}, T_{2,n},\ldots, T_{n,n}$ such that
\begin{align}
T \stackrel{d}{=} T_{1,n} + T_{2,n} + \ldots + T_{n,n},
\end{align}
where $\stackrel{d}{=}$ is equality in distribution.  
\end{definition}

The Gaussian, Poisson, negative binomial, gamma and Cauchy distributions are all infinitely divisible distributions on $\Re$. 

From the definition of a L\'evy process we see that for any $t>0$, $Y_t$ is a random variable belonging to the class of infinitely divisible distributions. Indeed, it follows from the fact that for any $n = 1,2,\ldots$, 
\begin{align}
Y_t = Y_{t/n} + (Y_{2t/n} - Y_{t/n}) + \ldots + (Y_t - Y_{(n-1)t/n})
\end{align}
together with the fact that $\{Y_t\}$ has stationary independent increments. 

The following lemma relates the characteristic exponent of $Y_t$ with that of $Y_1$. 
\begin{lemma}\cite[Chap. 2.1.]{Kuchler--Sorensen1997}\label{lemma.levycumulant}
For a L\'evy process $Y_t$, if $\mathbb{E} e^{i\theta Y_t} = e^{ \Psi_t(\theta)}$, then $\Psi_t(\theta) = t \Psi_1(\theta)$. 
\end{lemma}

Indeed, for two positive integers we have
\begin{align}
m \Psi_1(\theta)  = \Psi_m(\theta) = n \Psi_{m/n}(\theta),
\end{align}
which proves the statement for all rational $t>0$. The irrational cases follows from taking a limit and applying the right continuity of $X_t$ and the dominated convergence theorem. 

The full extent to which we may characterize infinitely divisible distributions is described by the L\'evy--Khintchine formula. 
\begin{lemma}[L\'evy--Khintchine formula]\cite{Sato1999}\label{lemma.lemmakhintchine}
A real-valued random variable $Y$ is infinitely divisible with characteristic function represented as
\begin{align}
\mathbb{E} e^{i \theta Y} & = e^{\Psi(\theta)},\quad \theta \in \mathbb{R},
\end{align}
if and only if there exists a triple $(a, \sigma, \nu)$, where $a\in \mathbb{R}, \sigma\geq 0$, and $\nu(\cdot)$ is a measure concentrated on $\mathbb{R} \backslash \{0\}$ satisfying $\int_{\mathbb{R}} (1 \wedge x^2) \nu(dx) <\infty$, such that
\begin{align}
\Psi(\theta) & = i a \theta - \frac{1}{2} \sigma^2 \theta^2 + \int_{\mathbb{R}} (e^{ i \theta z}-1 - i\theta z \mathbbm{1}_{|z|<1})\nu(dz). 
\end{align}
Moreover, for $\theta \in \{ \theta: \mathbb{E} e^{\theta Y}<\infty,  \theta \in \mathbb{R}\}$, we have the cumulant generating function of $Y$ as
\begin{align}\label{eqn.cdfnewrepre}
\kappa(\theta) & = a \theta + \frac{1}{2} \sigma^2 \theta^2 + \int_{\mathbb{R}} (e^{ \theta z}-1 - \theta z \mathbbm{1}_{|z|<1})\nu(dz). 
\end{align}
\end{lemma}

We call the tuple $(a,\sigma,\nu(dz))$ \emph{L\'evy characteristics} of the L\'evy process $\{Y_t\}$ if the characteristic function of $Y_1$ follows the L\'evy--Khintchine formula with triplet $(a,\sigma,\nu(dz))$. Particularly, we call the number $\sigma$ \emph{diffusion coefficient}, and the measure $\nu(dz)$ the \emph{L\'evy measure} of the L\'evy process $\{Y_t\}$. 

We have seen so far, that every L\'evy process can be associated with the law of an infinitely divisible distribution. The opposite, i.e. that given any random variable $X$, whose law of infinitely divisible, we can construct a L\'evy process $\{Y_t\}$ such that $Y_1 \stackrel{d}{=}X$. This is the subject of the L\'evy--It$\hat{\mathrm{o}}$ decomposition. 
\begin{lemma}\cite[Chap. 4]{Sato1999}[L\'evy--It$\hat{\mathrm{o}}$ decomposition]\label{lemma.levyitodecomposition}
Consider a triplet $(a,\sigma,\nu)$ where $a\in \mathbb{R}, \sigma\geq 0$ and $\nu$ is a measure satisfying $\nu(\{0\}) = 0$ and $\int_{\mathbb{R}} (1 \wedge x^2) \nu(dx)<\infty$. Then, there exists a probability space $(\Omega, \mathcal{F}, \mathbb{P})$ on which a L\'evy process $\{Y_t\}$ exists and decomposes as four independent processes as
\begin{align}\label{eqn.levyito}
Y_t = at + \sigma W_t + \int_0^t \int_{|z|< 1} z (\mu(ds,dz) - \nu(dz)ds) + \int_0^t \int_{|z| \geq 1} z \mu(ds,dz),
\end{align}
where $W_t$ is a standard Brownian motion, $\int_0^t \int_{|z|< 1} z (\mu(ds,dz) - \nu(dz)ds)$ is a square integrable pure jump martingale with an almost surely countable number of jumps of magnitude less than one on each finite time interval, and $\int_0^t \int_{|z| \geq 1} z \mu(ds,dz)$ is a compound Poisson process. The $\mu(dt,dz)$ is a jump measure defined to satisfy the following relations: $\forall \, \Gamma\in \mathcal{B}(\mathbb{R} \backslash \{0\})$,
\begin{equation}\label{eqn.jumpmeasuredef}
\mu((0,t] \times \Gamma) = \sum_{0<s\leq t} \mathbb{I}(\Delta Y_s \in \Gamma), 
\end{equation}
where $ \Delta Y_s = Y_s - Y_{s-}, Y_{s-} = \lim_{u\to s-} Y_u$. The measure $\nu(dz)$ is defined such that
\begin{equation}
\int_0^t \int_{|z|< 1} z (d\mu - \nu(dz)ds)
\end{equation}
is a martingale indexed by $t$. The measure $\nu(dz)ds$ is called the compensator for the multivariate point process $\mu(ds,dz)$. 

Moreover, the process $\{Y_t\}$ satisfies that
\begin{align}
\ln \mathbb{E} e^{i\theta Y_1} & = i a \theta - \frac{1}{2} \sigma^2 \theta^2 + \int_{\mathbb{R}} (e^{ i \theta z}-1 - i\theta z \mathbbm{1}_{|z|<1})\nu(dz). 
\end{align}
\end{lemma}

\subsection{Natural exponential family}\label{subsec.nef}

We briefly recall some notation and elementary properties of natural exponential families, which are widely used in statistics and probability~\cite{Kuchler--Sorensen1997}. 

Suppose $\mu$ is a probability measure on $\mathbb{R}$ with cumulant generating function $\kappa_\mu(\theta)$ defined as
\begin{align}
\kappa_\mu(\theta) & = \ln \int_{\mathbb{R}} e^{\theta y} \mu(dy).
\end{align}
We assume that the domain of $\theta$, $\Theta(\mu)=\{\theta: \kappa_\mu(\theta)<\infty\}$ is not empty. 

\begin{definition}[Natural exponential family]\label{def.nef}
The family of distributions given by
\begin{align}\label{eqn.exponentialfamilydefine}
P_{\theta,\mu}(dy) = e^{\theta y - \kappa_\mu(\theta)} \mu(dy)
\end{align}
is called the \emph{natural exponential family} generated by $\mu$. 
\end{definition}
If the measure $\mu$ follows an infinitely divisible distribution, so does all the distribution $P_{\theta, \mu}$. We omit the dependence on $\mu$ in $P_{\theta,\mu}(dy), \kappa_\mu(\theta), \Theta(\mu)$ and other quantities when the measure $\mu$ used is evident from context. 

The function $\kappa_\mu(\theta)$ is strictly convex and real analytic on $\Theta(\mu)$. We define the Fenchel--Legendre transform of the function $\kappa_\mu(\theta)$ as
\begin{align}\label{eqn.fenchellegendredefine}
\phi_\mu(x) & = \sup_{\theta \in \Theta(\mu)} (\theta x - \kappa_\mu(\theta)).
\end{align}

Denoting
\begin{align}
x(\theta) = \kappa_\mu'(\theta) = \int_{\mathbb{R}} y P_{\theta,\mu} (dy)
\end{align}
as the expectation of distribution $P_{\theta, \mu}$, it follows from convex duality that there exists an one-to-one function $\theta(x)$ such that
\begin{align}
\theta(x) & = \phi'_{\mu}(x(\theta)) 
\end{align}
Hence, we can index the distribution $P_{\theta,\mu}$ with either $\theta$ or $x$. The domain of $\theta$ is $\Theta(\mu)$, and the domain of $x$ is $M(P_{\theta,\mu}) =  \{x: x = \kappa'_\mu(\theta), \theta \in \Theta(\mu)\}$. 

To introduce the mean-value parametrized natural exponential family, we introduce the notion of the Bregman divergence below. 
\begin{definition}\label{def.bregmandivergence}
Let $f: \Omega \mapsto \mathbb{R}$ be a convex, continuously differentiable function, the domain $\Omega \subset \mathbb{R}^d$. Then, the Bregman divergence associated with $f$, denoted as $d_f(x,y)$, is defined as
\begin{align}
d_f(x,y) = f(x) - f(y) - \langle \nabla f(y), x-y \rangle,
\end{align}
where $\langle x,y \rangle$ denotes the inner product of $x$ and $y$. 
\end{definition}

It follows from Jensen's inequality that $d_f(x,y)\geq 0$. The Bregman divergence satisfies the following property when used as a loss function in Bayesian decision theory:
\begin{lemma}\label{lemma.meanpropertybregman}
Suppose $X$ is a random variable taking values in $\Omega$. Then, for any non-random element $u\in \Omega$, 
\begin{align}
\mathbb{E}[d_f(X,u)] & = \mathbb{E}[d_f(X, \mathbb{E}[X])] + d_f(\mathbb{E}[X],u),
\end{align}
where the expectations are taken with respect to the distribution of $X$. 
\end{lemma}

\begin{IEEEproof}
It follows from straightforward algebra that
\begin{align}
d_f(X,u) & = d_f(X, \mathbb{E}[X]) + d_f(\mathbb{E}[X],u) + \langle f'(\mathbb{E}[X]) - f'(u), X - \mathbb{E}[X] \rangle. 
\end{align}
Taking expectations on both sides finishes the proof. 
\end{IEEEproof}

It follows from Lemma~\ref{lemma.meanpropertybregman} that
\begin{align}
\mathbb{E}[X] & = \argmin_{u\in \Omega} \mathbb{E}[d_f(X,u)]. 
\end{align}
Further, if $f$ is strictly convex, then $\mathbb{E}[X]$ uniquely solves $\min_u \mathbb{E}[d_f(X,u)]$. 

The following well known lemma~(see, e.g.,~\cite{Nielsen--Nock2010}) characterizes the likelihood ratio and relative entropy between distributions from the same natural exponential family.
\begin{lemma}\label{lemma.regular}
Suppose $P_{\theta}$ is the natural exponential family in~(\ref{eqn.exponentialfamilydefine}), $\phi(x)$ is defined in~(\ref{eqn.fenchellegendredefine}), and $x(\theta) = \kappa'(\theta)$ is the mean parameter. Let $x_1 = x(\theta_1), x_2 = x(\theta_2)$. 

Then, 
\begin{align}
\frac{dP_{\theta_1}}{dP_{\theta_2}} & = e^{-d_{\phi}(y, x_1) + d_\phi(y,x_2)} \\
D(P_{\theta_1} \| P_{\theta_2}) & = \phi(x_1) - \phi(x_2) - \phi'(x_2)(x_1 - x_2), \label{eq:d-bregman}
\end{align}
where $d_\phi(x_1,x_2)$ is the Bregman divergence generated by convex function $\phi(\cdot)$. 
\end{lemma}

\begin{IEEEproof}
It follows from convex duality that
\begin{align}
\kappa(\theta(x)) & = \theta(x) x - \phi(x),
\end{align}
where $\theta(x) = \phi'(x)$. 

Then, it follows from the definition of $P_{\theta}$ that 
\begin{align}
\ln \frac{dP_{\theta_1}}{dP_{\theta_2}} & = \theta_1 y - \kappa(\theta_1) - (\theta_2 y - \kappa(\theta_2)) \\
& = \phi'(x_1) y - (\phi'(x_1)x_1 - \phi(x_1)) - \phi'(x_2)y + (\phi'(x_2) x_2 - \phi(x_2)) \\
& = \left( \phi(y) - \phi(x_2) - \phi'(x_2)(y - x_2) \right) - \left( \phi(y) - \phi(x_1) - \phi'(x_1)(y-x_1) \right) \\
& = d_\phi(y,x_2) - d_\phi(y,x_1). 
\end{align}

Taking expectation on both sides with respect to $P_{\theta_1}$, we have
\begin{align}
D(P_{\theta_1} \| P_{\theta_2}) & = \int \ln \frac{dP_{\theta_1}}{dP_{\theta_2}} dP_{\theta_1} \\
& = \phi'(x_1) x_1 - (\phi'(x_1)x_1 - \phi(x_1)) - \phi'(x_2)x_1 + (\phi'(x_2) x_2 - \phi(x_2)) \\
& = \phi(x_1) - \phi(x_2) - \phi'(x_2)(x_1 - x_2) \\ & = d_\phi(x_1,x_2). 
\end{align}
\end{IEEEproof}


%

\subsection{The Discrete-time L\'evy channel}

The discrete-time L\'evy channel is a special case of the natural exponential family.

\begin{definition}[Discrete-time L\'evy channel]\label{def.discrete-timelevy}
For a L\'evy process $Y_t$ with characteristic triplet $(a,\sigma,\nu)$, denote $\kappa(\theta) = \ln \mathbb{E} e^{\theta Y_1}$ and $\phi(x) = \sup_{\theta} (\theta x - \kappa(\theta))$. Let $\gamma>0$, and $\mu_\gamma$ be an infinitely divisible probability distribution with cumulant generating function $\gamma \cdot \kappa(\theta)$. Then, the discrete-time L\'evy channel generated by L\'evy characteristics $(a,\sigma,\nu)$ (or by cumulant generating function $\kappa(\theta)$) at SNR $\gamma$ is given by
\begin{align}\label{eq: c.o.m. levy channel}
P_{Y_\gamma|X} = P_{\phi'(X), \mu_\gamma},
\end{align} 
where $P_{\theta,\mu}$ is the natural exponential family in Definition~\ref{def.nef}. Concretely, for any Borel set $A\subset \mathbb{R}$, we have
\begin{align}
P_{Y_\gamma|X}(A) = \int_{z\in A} e^{\phi'(X)z - \gamma \kappa(\phi'(X))} \mu_\gamma(dz),
\end{align}
where the input $X$ lies in set $\{\kappa'(\theta): \int_{\R} \frac{z^2}{1+z^2}e^{\theta z}\nu(dz) <\infty \}$ that is assumed to have non-empty interior. 
\end{definition}

The input domain of $X$ is determined by the values of $\theta$ that make $P_{Y_\gamma|X}$ a valid infinitely divisible distribution (See~\cite[Eqn. 2.1.10.]{Kuchler--Sorensen1997}). We further emphasize that the domain of inputs having non-empty interior is necessary. For example, in the case of the Cauchy distribution of probability density function $\frac{1}{\pi (1+x^2)}, x \in \R$, we have $\sigma = 0, \nu(dz) = \frac{dz}{\pi z^2},z\in \R, z\neq 0$, whose cumulant generating function $\kappa(\theta)$ is not infinity only when $\theta = 0$. 

In short, the discrete-time L\'evy channel is the natural exponential family generated by a distribution with cumulant generating function $\gamma \cdot  \kappa(\theta)$, while $\theta = \phi'(X), X = \kappa'(\theta)$. We have the following lemma characterizing the distribution of $Y_\gamma$ conditioned on $\theta$ (or equivalently, $X$):
\begin{lemma}\label{lemma.outputcgf}
Suppose $P_{Y_\gamma|X}$ is the discrete-time L\'evy channel generating by cumulant generating function $\kappa(\theta)$ at SNR $\gamma$. Then, with the convention of $\theta = \phi'(X), X = \kappa'(\theta)$, 
\begin{align}
\ln \mathbb{E}_{Y_\gamma|X} e^{s Y_\gamma} & = \gamma (\kappa(\theta + s) - \kappa(\theta)) \\
& = \gamma ( \kappa(\phi'(X) + s) - \kappa(\phi'(X)) ). \label{eqn.cumulantofoutput}
\end{align}
\end{lemma}
\begin{IEEEproof}
We compute the cumulant generating function as follows:
\begin{align}
\ln \mathbb{E}_{Y_\gamma|X} e^{s Y_\gamma} & = \ln \int e^{sz} e^{\theta z - \gamma \kappa(\theta)} \mu_\gamma(dz) \\
& = \ln e^{ - \gamma \kappa(\theta)}  \int e^{(s+\theta) z} \mu_\gamma(dz) \\
& = -\gamma \kappa(\theta) + \gamma \kappa(s+\theta) \\
& = \gamma \left( \kappa(s+\theta) - \kappa(\theta) \right),
\end{align}
where we used the fact that $\mu_\gamma$ has cumulant generating function $\gamma \kappa(\theta)$. 
\end{IEEEproof}

The following lemma characterizes the change of the L\'evy characteristics while one varies the input to the discrete-time L\'evy channel. It suffices to consider the case of SNR equal to one. 
\begin{lemma}
Suppose $P_{Y_1|X}$ is the discrete-time L\'evy channel generated by L\'evy characteristics $(a,\sigma,\nu)$ at SNR $\gamma =1$. Then, with the convention that $\theta = \phi'(X), X = \kappa'(\theta)$, the L\'evy characteristics of the output conditioned on $\theta$ is as follows:
\begin{align}
a_\theta & = a + \sigma^2 \theta + \int_{\mathbb{R}} x \mathbbm{1}_{|x|<1} (e^{\theta x}-1) \nu(dx)\\
\sigma_\theta & = \sigma \\
\nu_\theta(dz) & = e^{\theta z} \nu(dz)
\end{align}
\end{lemma}

\begin{IEEEproof}
It follows from the L\'evy--Khintchine formula that the cumulant generating function of $Y_1$ admits the expression
\begin{align}
\kappa(s) & = a s + \frac{1}{2} \sigma^2 s^2 + \int_{\mathbb{R}} (e^{ s z}-1 -s z \mathbbm{1}_{|z|<1})\nu(dz). 
\end{align}

It follows from Lemma~\ref{lemma.outputcgf} that
\begin{align}
\kappa_\theta(s) & = \kappa(s+\theta) - \kappa(\theta) \\
& = as + \frac{1}{2}\sigma^2 s^2 + \sigma^2 s \theta + s \int_{\mathbb{R}} z \mathbbm{1}_{|z|<1} (e^{\theta z}-1) \nu(dz) + \int_{\R} (e^{sz} -1 - sz \mathbbm{1}_{|z|<1})e^{\theta z}\nu(dz),
\end{align}
which implies the claimed result. 
\end{IEEEproof}

We now shed some light on the parameter $\gamma$ which is an integral part of our characterization of the L\'evy channel above. An important feature of the channel in Definition \ref{def.discrete-timelevy} is that it endows $\gamma$ with a very natural interpretation as the ``SNR level'' for the L\'evy channel. It turns out, that we can place the different $\{Y_\gamma\}$ on the same probability space to construct a L\'evy process $\{Y_\gamma\}_{0 \leq \gamma \leq T}$ indexed by $\gamma$, such that the marginal distribution of $Y_\gamma$ follows the distribution specified in Definition~\ref{def.discrete-timelevy} for every $\gamma \in [0,T]$. 

We construct the coupling as follows. For the L\'evy process $\{Y_t\}$ that satisfies $\ln \mathbb{E} e^{\theta Y_1} = \kappa(\theta)$, it follows from Lemma~\ref{lemma.levycumulant} that $\ln \mathbb{E} e^{\theta Y_t} = t \kappa(\theta)$. We have the following result.
\begin{lemma}\cite[Chap. 2]{Kuchler--Sorensen1997}\label{lemma.rdderivative}
Let $P_\theta^{[0,T]}$ denote the probability measure for the L\'evy process $\{Z_t, t\in [0,T]\}$ restricted to the natural filtration $\mathcal{F}_t^Z = \sigma\{Z_s: 0\leq s\leq t\}$ that satisfies $\ln \mathbb{E}_{P_\theta^{[0,T]}} e^{s Z_1} = \kappa(s+\theta) - \kappa(\theta)$, and $P_0^{[0,T]}$ denote the measure when $\theta = 0$. Then, $P_\theta^{[0,T]} \ll P_0^{[0,T]}$, and the Radon--Nikodym derivative can be expressed as
\begin{align}\label{eqn.changeofmeasurewhole}
\frac{dP_\theta^{[0,T]}}{dP_0^{[0,T]}} = e^{\theta Z_T - T \kappa(\theta)}. 
\end{align}
\end{lemma}

It is clear that the marginal distribution of $Z_t$ under $P_\theta^{[0,T]}$ is equal to the distribution of $Y_\gamma$ in Definition~\ref{def.discrete-timelevy} when we set $t = \gamma$ and $\theta = \phi'(X)$. Hence, for a fixed $\theta$ (or equivalently, $X$), one can view the output of the discrete-time L\'evy channel, $Y_\gamma$, as the value of the random process $Z_t$ at time $t = \gamma$. 

In this context, (\ref{eqn.changeofmeasurewhole}) shows that for a statistical problem where the goal is to infer $\theta$ (or equivalently, $X$) from observing the whole process $\{Z_t: 0\leq t\leq T\}$, the final observation $Z_T$ is the sufficient statistic. In other words, we have a natural degradedness in the observations, in terms of the index $\gamma$. This in our opinion gives the parameter $\gamma$ the most suitable interpretation as the signal-to-noise ratio of the channel.

Before we proceed, it would be instructive to understand how the Gaussian and Poisson channels are subsumed in our framework of L\'evy channels. 

\begin{example}[Gaussian channel]\label{example.gaussian}
Let $\kappa(\theta) = \frac{1}{2}\theta^2$, which corresponds to $\mathcal{N}(0,1)$. We have $X = \kappa'(\theta) = \theta$. It follows from (\ref{eqn.cumulantofoutput}) that the output random variable $Y_\gamma|X$ has cumulant generating function
\begin{align}
\gamma \cdot (\kappa(X + s) - \kappa(X)) & = \gamma X s + \frac{\gamma s^2}{2},
\end{align}
which corresponds to distribution $\mathcal{N}(\gamma X, \gamma)$, recovering the definition of the Gaussian channel at SNR $\gamma$.
\end{example}

\begin{example}[Poisson channel]\label{example.poisson}
Let $\kappa(\theta) = e^{\theta}-1$, which corresponds to $\mathsf{Poi}(1)$. We have $X = \kappa'(\theta) = e^\theta$, hence $\theta = \ln X$. It follows from (\ref{eqn.cumulantofoutput}) that the output random variable $Y_\gamma|X$ has cumulant generating function
\begin{align}
\gamma \cdot (\kappa( \ln X + s) - \kappa( \ln X )) & = \gamma X(e^s -1),
\end{align}
which corresponds to distribution $\mathsf{Poi}(\gamma X)$, recovering the definition of the Poisson channel at SNR $\gamma$. 
\end{example}

\section{Main results} \label{sec: main results}
\begin{quote}
{``It is even speculated (in \cite{Guo--Shamai--Verdu2005}) that information and estimation satisfy similar relationships as long as the output has independent increments conditioned on the input.''}  \flushright{-- Guo, Shamai and Verd\'u \cite{Guo--Shamai--Verdu2008}}
\end{quote}
To some extent, our work gives a clear affirmative answer to the above suspicion raised in the context of the Gaussian and Poisson results from Section \ref{sec: intro}. In this section, we will observe that analogous to the Gaussian and Poisson channel, we are able to obtain, for the L\'evy channel, a precise formula expressing the mutual information as an optimal estimation loss. As we will shortly demonstrate, the ``correct'' loss function that presents itself in this formula, is intimately connected with the Gaussian and Poisson loss functions that we visited in Section \ref{sec: intro}.

Formally, let $\Xscr \subset \mathbb{R}$ denote the space of channel inputs as defined in Definition~\ref{def.discrete-timelevy}. The reconstruction space for estimating the channel input $x \in \Xscr$, denoted by $\hat{\Xscr}$ is a function space. Each reconstruction $\hat{x} \in \hat{\Xscr}$ is a collection of scalars indexed by $\Re$. In other words, $\hat{x} = \{\hat{x}_z : z \in \Re, \hat{x}_z\geq 0\}$. We will now introduce the loss function for the L\'evy channel. 

\begin{definition}[Loss function for discrete-time L\'evy Channels] \label{def:levy-channel-loss}
The loss function $\ell_{\Lscr}: \Xscr \times \hat{\Xscr} \to [0, \infty]$, for the L\'evy channel with characteristics $(a,\sigma, \nu(dz))$, is defined as, 
\begin{align} \label{eq:levy-loss}
\ell_{\Lscr}(x, \hat{x}) \doteq \sigma^2 \ell_{\Gscr} (\phi'(x), \hat{x}_0) + \int_{\R} \ell_{\Pscr} (e^{\phi'(x)z}, \hat{x}_z)\, \nu(dz),
\end{align} 
where the loss functions $\ell_{\Gscr}$ (\ref{eq: gaussian loss}) and $\ell_{\Pscr}$ (\ref{eq: poisson loss}) are as defined in Section \ref{sec: intro}.
\end{definition}  
The loss function for L\'evy channels has some interesting properties. It is always non-negative, and achieves zero if and only if 
\begin{equation}
\hat{x}_0 = \phi'(x), \hat{x}_z = e^{\phi'(x)z}, z \neq 0, \nu \text{-a.s.}
\end{equation}

The reconstruction can be viewed as performing an individual estimate for every jump size $z$ for the pure jump part of the L\'evy process indexed by SNR level, in addition to a single estimate for the continuous part of the channel output. Applying Lemma~\ref{lemma.meanpropertybregman} to the Bregman divergence $\ell_{\Lscr}(x, \hat{x})$, we have the following result. 
\begin{lemma}\label{lemma.bregmanwhole}
Suppose $X$ is a random variable on probability space $(\Omega, \mathcal{F}, \mathbb{P})$. Then, for any $\mathcal{H}$-measurable reconstruction $\hat{x} \in \hat{\mathcal{X}}$, where $\mathcal{H} \subset \mathcal{F}$, we have
\begin{align}
\mathbb{E}[\ell_{\Lscr}(X, \hat{x})|\mathcal{H}] & = \sigma^2 \ell_{\Gscr}(\phi'(X), \mathbb{E}[\phi'(X)|\mathcal{H}]) + \int_{\R} \ell_{\Pscr}(e^{\phi'(X)z}, \mathbb{E}[e^{\phi'(X)z}|\mathcal{H}])\nu(dz) \\
& \quad + \sigma^2 \ell_{\Gscr}(\mathbb{E}[\phi'(X)|\mathcal{H}], \hat{x}_0) + \int_{\R} \ell_{\Pscr}(\mathbb{E}[e^{\phi'(X)z}|\mathcal{H}], \hat{x}_z)\nu(dz) \\
\end{align}
\end{lemma}

Lemma~\ref{lemma.bregmanwhole} shows that the unique $\mathcal{H}$-measurable reconstructions that minimizes the conditional Bayes risk is the conditional expectations (optimal reconstructions) for each jump size $z$, separately. We are now in a position to present our first main result for information and estimation in the L\'evy channel, which presents a formula for the mutual information between the input and output of the L\'evy channel. 


\begin{theorem} \label{thm:levy-i-mmle}
Let $X$ be a real-valued random variable distributed according to law $P$. For the discrete-time L\'evy channel generated by L\'evy characteristics $(a,\sigma,\nu)$, suppose the following is true:
\begin{enumerate}
\item If $\sigma\neq 0$, then $\mathbb{E}_P (\phi'(X))^2 <\infty$.
\item If $\int_{\R} \nu(dz)<\infty$, then $\mathbb{E}_P \int_{\R} |\phi'(X)z|e^{\phi'(X)z}\nu(dz)<\infty$. 
\item If $\int_{\R} \nu(dz) = \infty$, then $\mathbb{E}_P \int_{\R} |\phi'(X)z|e^{\phi'(X)z}\nu(dz)<\infty$, and for any $0<\gamma<\infty$,
\begin{align}
\int_0^\gamma \int_{\mathbb{R}} \mathbb{E}   \left( e^{\phi'(X)z} \Bigg|\ln \E_P[e^{\phi'(X)z}|Y_\alpha]\Bigg| \right) \nu(dz)d\alpha & <\infty. 
\end{align} 
\end{enumerate}

Then, 
\begin{align} \label{eq:levy-i-mmle}
\frac{\partial}{\partial \gamma} I(X; Y_\gamma) =  \E[ \ell_{\Lscr} (X, \hat{X}^P_\gamma) ] ,
\end{align}
where $\hat{X}^P_\gamma$, defined as,
\vspace{-2pt}
\begin{align}
\hat{X}^P_{\gamma,z} = \twopartdef{\E_P[\phi'(X)| Y_\gamma]}{z = 0,}{\E_P[e^{\phi'(X)z} | Y_\gamma]}{ z \neq 0,}
\end{align}
is the optimal (minimum mean loss) reconstruction as shown in Lemma~\ref{lemma.bregmanwhole}. 
\end{theorem}

It is evident from Theorem~\ref{thm:levy-i-mmle} that only for the Gaussian and Poisson channels with a single jump size can the reconstructions be reduced to a single estimator. The result in (\ref{eq:levy-i-mmle}) presents the derivative of the mutual information between the input and output with respect to the SNR as the optimal mean loss in estimating the channel input, according to the loss function specified in (\ref{eq:levy-loss}). It is strikingly similar to the I-MMLE results encountered in Section \ref{sec: intro} for the Gaussian (\ref{eq: i-mmse}) and Poisson (\ref{eq: poisson i-mmle}) channels. In fact, in the next section we will demonstrate that Theorem \ref{thm:levy-i-mmle} directly implies both these results. We mention that for certain special cases, such as the Gaussian setting, the assumptions can be weakened to the condition that the mutual information $I(X; Y_\gamma)<\infty$ for some $\gamma>0$, as shown in~\cite[Thm. 6]{wu2012functional}. 

A particularly interesting case arises when $\gamma = \infty$, since it gives us a new expression for the entropy of a random variable. 

\begin{theorem} \label{thm:h-mmle}
If $X$ is a discrete real-valued random variable. For the discrete-time L\'evy channel generated by L\'evy characteristics $(a,\sigma,\nu)$, suppose the following is true:
\begin{enumerate}
\item If $\sigma\neq 0$, then $\mathbb{E}_P (\phi'(X))^2 <\infty$.
\item If $\int_{\R} \nu(dz)<\infty$, then $\mathbb{E}_P \int_{\R} |\phi'(X)z|e^{\phi'(X)z}\nu(dz)<\infty$. 
\item If $\int_{\R} \nu(dz) = \infty$, then $\mathbb{E}_P \int_{\R} |\phi'(X)z|e^{\phi'(X)z}\nu(dz)<\infty$, and for any $0<\gamma<\infty$,
\begin{align}
\int_0^\gamma \left( \int_{\mathbb{R}} \mathbb{E}   \left( e^{\phi'(X)z} \Bigg|\ln \E_P[e^{\phi'(X)z}|Y_\alpha]\Bigg| \right) \nu(dz) \right) d\alpha & <\infty. 
\end{align} 
\end{enumerate}
Then, 
\begin{equation} \label{eq:h-mmle}
H(X) = \int_0^\infty \E[ \ell_{\Lscr} (X, \hat{X}^P_\gamma) ] \, d \gamma.
\end{equation}
\end{theorem}
The above representation of entropy in Theorem \ref{thm:h-mmle} is quite intriguing. In particular, it holds for any L\'evy channel. Further, the left hand side, as is well known, is invariant to one-to-one transformations, a fact that is not at all intuitive for the right hand side. Indeed, the fact that such a functional representation for the entropy of a random variable holds in general, is surprising. 

We will now visit the mismatched estimation setting, and present a result analogous to Theorem \ref{thm:levy-i-mmle} in this direction. Recall that in the case of the mismatched decoder, the true law governing the channel input is $P$ while the decoder incorrectly believes it to be $Q$. It thus employs the estimator optimized for $Q$. Using a sub-optimal decoder will incur an additional loss. This loss is also termed ``cost of mismatch''. In the following theorem, we demonstrate that two quantities, namely the relative entropy between the true and mismatched channel output laws, and the integral with respect to SNR of the ``cost of mismatch'' - are exactly equal. 
\begin{theorem} \label{thm:D-mle}
Let $X \in \mathcal{X}$ be a real-valued random variable, and $P$ and $Q$ be two laws on $\mathcal{X}$. For the discrete-time L\'evy channel generated by L\'evy characteristics $(a,\sigma,\nu)$, suppose the following is true at SNR $\gamma>0$:
\begin{enumerate}
\item If $\sigma \neq 0$, then
\begin{align} \label{eqn.maintheoremcond1}
\E_P \int_0^\gamma \left( \E_P[\phi'(X)|Y_\alpha] - \E_Q[\phi'(X)|Y_\alpha] \right)^2 d\alpha & <\infty
\end{align}
\item \begin{align}\label{eqn.maintheoremcond2}
\int_0^\gamma \left( \int_{\mathbb{R}} \E_P \left( \E_P[e^{\phi'(X)z}|Y_\alpha] \Bigg|\ln \left( \frac{\E_P[e^{\phi'(X)z}|Y_\alpha]}{\E_Q[e^{\phi'(X)z}|Y_\alpha]} \right)\Bigg| \right) \nu(dz) \right) d\alpha & <\infty
\end{align}
\end{enumerate}
Then, 
\begin{align} \label{eq:levy-d-mle}
D(P_{Y_\gamma} || Q_{Y_\gamma}) = \int_0^\gamma \E_{P}[ \ell_{\Lscr} (X, \hat{X}^Q_\alpha)  - \ell_{\Lscr} (X, \hat{X}^P_\alpha)] \, d\alpha,
\end{align}
where $P_{Y_\gamma}$ ($Q_{Y_\gamma}$) denotes the law of the channel output at SNR level $\gamma$, when the true (mismatched) input law is $P$ ($Q$).  
\end{theorem}   

When we consider the case $\gamma = \infty$, we obtain the following representation of relative entropy:
\begin{theorem}\label{thm.levyrelativerepre}
Let $X \in \mathcal{X}$ be a real-valued random variable, and $P$ and $Q$ be two laws on $\mathcal{X}$. For the discrete-time L\'evy channel generated by L\'evy characteristics $(a,\sigma,\nu)$, suppose the conditions in~(\ref{eqn.maintheoremcond1}) and~(\ref{eqn.maintheoremcond2}) are true for all $0< \gamma <\infty$. 
Then,
\begin{equation}
D(P \| Q) = \int_0^\infty \E_{P}[ \ell_{\Lscr} (X, \hat{X}^Q_\gamma)  - \ell_{\Lscr} (X, \hat{X}^P_\gamma)] \, d\gamma,
\end{equation}
\end{theorem}


\section{Recovering Gaussian and Poisson results} \label{sec: recover G and P}

\subsection{Gaussian channel}
As we showed in Example~\ref{example.gaussian}, the discrete-time L\'evy channel specialized to the Gaussian setting corresponds to the channel 
\begin{align} \label{eq: discrete-time gaussian channel}
Y_\gamma = \gamma X + W,
\end{align}
where $W$ is a Gaussian random variable $\mathcal{N}(0,\gamma)$ independent of $X$. We have $\kappa(\theta) = \frac{1}{2}\theta^2, X = \kappa'(\theta)$, and the L\'evy characteristics are $a = 0, \sigma = 1, \nu(dz) \equiv 0$. 

Under this framework, it is easy to see that the loss function in (\ref{def:levy-channel-loss}) collapses to $\ell_\Gscr$, i.e., the squared error loss function defined in (\ref{eq: gaussian loss}). Also recall, from Section \ref{sec: levy-type channels}, that for the Gaussian channel we have $\theta = X, \kappa(\theta) = \frac{1}{2}\theta^2, \phi(X) = \frac{1}{2}X^2$. The assumptions reduce to $\E[X^2] < \infty$, which is consistent with classical results, cf. \cite[Theorem 1]{Guo--Shamai--Verdu2005}. Further, an application of Theorem \ref{thm:levy-i-mmle} directly gives us the I-MMSE formula \cite{Guo--Shamai--Verdu2005}, as stated in (\ref{eq: i-mmse})-(\ref{eq: gaussian loss}),
\begin{align}
\frac{\partial}{\partial \gamma} I(X; Y_\gamma) = \E[ \ell_{\Gscr} (X, \E[X|Y_\gamma]) ].
\end{align} 
By an essentially identical argument, we know from Theorem~\ref{thm.levyrelativerepre} that if $\int_0^\gamma \E_P(\E_{P}[X|Y_\alpha] - \E_{Q}[X|Y_\alpha])^2d \alpha<\infty$ for all $\gamma>0$, we have
\begin{align}
D( P \| Q) = \int_0^{\infty} \E_{P}[ \ell_{\Gscr} (X, \E_{Q}[X|Y_\alpha]) - \ell_{\Gscr} (X, \E_{P}[X|Y_\alpha]) ] \,d\alpha,
\end{align}
which is Verd{\'u}'s result in \cite{Verdu2010}.

\subsection{Poisson channel}

As we showed in Example~\ref{example.poisson}, the discrete-time L\'evy channel specialized to the Poisson setting corresponds to the channel 
\begin{align}
Y_\gamma|X \sim \mathsf{Poi}(\gamma X). 
\end{align}
We have $\kappa(\theta) = e^\theta -1, X = e^\theta, \theta = \ln X$. The L\'evy characteristics are $a = 0, \sigma = 0, \nu(dz) = \delta_1$, where the measure $\delta_x$ denotes a point mass at $x$. 

Again, it is straightforward to see that the loss function in (\ref{eq:levy-loss}) collapses to the natural Poisson loss function $\ell_\Pscr$ defined in (\ref{eq: poisson loss}). The assumptions reduce to $\E[X \ln X] < \infty$, a condition identical to the treatment in the literature, cf. \cite[Section V.A]{Atar--Weissman2012}. An application of Theorem \ref{thm:levy-i-mmle} to this specialized setting, gives us the relationship between mutual information and minimum mean loss in estimation for the Poisson channel,
\begin{align}
\frac{\partial}{\partial \gamma} I(X; Y_\gamma) = \E [ \ell_{\Pscr}(X, \E[X|Y_\gamma]) ].
\end{align}

Applying Theorem~\ref{thm.levyrelativerepre} to the Poisson channel, we obtain the following corollary:
\begin{corollary}\label{cor.poidiscrete-time}
Let non-negative random variable $X \in \mathcal{X}$ be the input to the Poisson channel, $P,Q$ are two probability measures on $\mathcal{X}$. Suppose for all $0< \gamma <\infty$, 
\begin{align}
\int_0^\gamma \mathbb{E}_P \left( \E_{P}[X|Y_\alpha] \left | \ln \frac{\E_{P}[X|Y_\alpha]}{\E_{Q}[X|Y_\alpha]} \right |  \right) d\alpha <\infty. 
\end{align}
Then, 
\begin{align}
D( P \| Q ) = \int_0^{\infty} \E_{P}[ \ell_{\Pscr} (X, \E_{Q}[X|Y_\alpha]) - \ell_{\Pscr} (X, \E_{P}[X|Y_\alpha]) ] \,d\alpha. 
\end{align} 
\end{corollary}

It is worth mentioning that Corollary~\ref{cor.poidiscrete-time} is a strengthened version of Atar and Weissman\cite[Thm 4.1]{Atar--Weissman2012}. In \cite[Thm 4.1]{Atar--Weissman2012} and \cite{Guo--Shamai--Verdu2013}, the mismatched estimation results are stated with the very strong condition that $P$ and $Q$ are probability measures supported on interval $[a,b], 0<a<b<\infty$. This condition appears to be restrictive to the authors, since it eliminates the possibility of using zero input in the Poisson channels. Indeed, it was conjectured in \cite{Atar--Weissman2012} that their Theorem 4.1 holds under weaker assumptions, which Corollary~\ref{cor.poidiscrete-time} presents. 


\section{Two new examples: the gamma and negative binomial channels} \label{sec: gamma}

\subsection{gamma channel}


Recall that the gamma distribution is a continuous-valued two-parameter probability distribution. We say that a random variable $Z$ follows the gamma distribution with ``shape'' parameter $k>0$ and ``scale'' parameter $\alpha>0$, or equivalently $Z \sim \Gamma(k,\alpha)$, if it has the following probability density function:
\begin{align} \label{eq:gamma-distribution}
f(z;k,\alpha) = \frac{z^{k-1} e^{-z/\alpha}}{\alpha^k \Gamma(k)}, z>0.
\end{align}
The gamma distribution satisfies infinite divisibility with respect to the shape parameter, for a fixed scale. The cumulant generating function of $Z\sim \Gamma(k,\alpha)$ is $k \ln \left( \frac{1}{1-\alpha\theta} \right), \theta <\frac{1}{\alpha}$.

%
%
\begin{definition}[gamma channel]
Set $\kappa_\Gamma(\theta) = \ln \left( \frac{1}{1 - \theta} \right), \theta<1$. Then, the output of the discrete-time L\'evy channel generated by $\kappa_\Gamma(\theta)$ follows
\begin{align} \label{eq:gamma-channel}
Y_\gamma |X \sim \Gamma (\gamma, X),\quad X>0. 
\end{align}
\end{definition}

Indeed, we have $X = \kappa_\Gamma'(\theta) = \frac{1}{1-\theta}, \theta = 1-\frac{1}{X}$. It follows from Lemma~\ref{lemma.outputcgf} that the cumulant generating function of the output is
\begin{align}
\ln \mathbb{E}_{Y_\gamma|X} e^{sY_\gamma} & = \gamma \left ( \ln \left( \frac{1}{1-s-\theta} \right) - \ln \left( \frac{1}{1-\theta} \right) \right ) \\
& = \gamma \ln \left( \frac{1}{1-Xs} \right),
\end{align}
which implies that $Y_\gamma|X \sim \Gamma(\gamma,X)$. 

Now we compute the L\'evy characteristics corresponding to $\kappa_\Gamma(\theta) = \ln \left( \frac{1}{1-\theta} \right)$. We have
\begin{align}
\kappa_\Gamma'(\theta) & = \frac{1}{1-\theta} \\
& = \int_{0}^\infty e^{-(1-\theta)z}dz. 
\end{align}
Integrating on both sides with respect to $\theta$ and utilizing $\kappa_\Gamma(0) = 0$, we have
\begin{align}
\kappa_\Gamma(\theta) & = \int_0^\infty \frac{e^{\theta z }-1}{z} e^{-z}dz \\
& = \int_0^\infty (e^{\theta z} -1) \nu(dz),
\end{align}
where $\nu(dz) = e^{-z} z^{-1}dz, z>0$. Note that $\int_{\R} \nu(dz) = \infty$. 

Hence, we know that for the gamma channel, $\sigma = 0, \nu(dz) = e^{-z} z^{-1}dz, z>0$. The channel defined in (\ref{eq:gamma-channel}) is a very simple observation model. The input simply modulates the scale parameter of a gamma distributed random variable. It is therefore striking that this probabilistic model joins the elite group of channels which enjoy a unique relationship between mutual information and optimal estimation. We summarize this in the following result, a direct consequence of Theorem \ref{thm:levy-i-mmle}. 
\begin{theorem} \label{thm:gamma-i-mmle}
Let non-negative random variable $X \sim P$ be the input to the gamma channel that satisfies $\mathbb{E}X<\infty$ and for any $0<\gamma<\infty$,
\begin{align}
\int_0^\gamma \left( \int_{z\geq 0} \mathbb{E} z^{-1}  \left( e^{-z/X} \Bigg|\ln \E_P[e^{z-z/X}|Y_\alpha]\Bigg| \right) dz \right) d\alpha & <\infty. 
\end{align} 
Let $Y_\gamma$ denote the channel output at SNR $\gamma$. Then, we have,
\begin{align}
\frac{\partial}{\partial \gamma} I(X; Y_\gamma) = \E[ \ell_{\Gamma} (X, \hat{X}^P_\gamma) ],
\end{align}
where the loss function $\ell_{\Gamma}$ is defined as,
\begin{align} \label{eq:gamma-loss}
\ell_\Gamma (x, \hat{x}) = \int_{\Re^{+}} \ell_\Pscr (e^{(1 - 1/x)z}, \hat{x}_z) z^{-1} e^{-z}\, dz 
\end{align}
and the reconstruction $\hat{X}^P_\gamma = \{\hat{X}^P_{\gamma,z}: z>0\} \in \mathcal{X}$ satisfies
\begin{align}\label{eqn.gammaestimatorsdef}
\hat{X}^P_{\gamma,z} = \mathbb{E}[e^{(1-1/X)z}|Y_\gamma],\quad \forall z>0.
\end{align}
Recall that $\ell_\Pscr$ is the Poisson loss function introduced in (\ref{eq: poisson loss}).
\end{theorem}
Theorem \ref{thm:gamma-i-mmle} can also be extended to incorporate mismatch. We now state this extension in the following result.
\begin{theorem} \label{thm:gamma-dmle}  
Let $X$ be the input to the gamma channel, and conditions in Theorem~\ref{thm:D-mle} are satisfied. Let $P_{Y_\gamma}$ and $Q_{Y_\gamma}$ denote the output laws when the input is distributed according to $P$ and $Q$ respectively. Then, we have,
\begin{align}
D(P_{Y_\gamma} || Q_{Y_\gamma}) = \int_0^\gamma \E_P[ \ell_{\Gamma} (X, \hat{X}^Q_\alpha) - \ell_{\Gamma} (X, \hat{X}^P_\alpha) ]\, d\alpha,
\end{align}
The loss function $\ell_\Gamma$ is defined in (\ref{eq:gamma-loss}), and the reconstructions $\hat{X}_\alpha^{P}$ and $\hat{X}_\alpha^Q$ are defined in~(\ref{eqn.gammaestimatorsdef}). 
\end{theorem}



\subsection{negative binomial channel}

We now turn our attention to another interesting example from the family of discrete-time L\'evy channels - the negative binomial channel. Recall that the negative binomial distribution is a discrete law, that governs the number of independent Bernoulli trials required to obtain a specified number of failures. Specifically, we say that $Z$ is distributed according to the negative binomial distribution with parameters $r$ and $p$, or equivalently, $Z \sim \mathsf{NB}(r, p)$, with $r > 0, 0 \leq p \leq 1$, if it has the probability mass function,
\begin{align}
p(k; r, p) = \frac{1}{k!}\,\frac{\Gamma(k+r)}{\Gamma(r)} (1-p)^r p^k , k = 0, 1, 2 \ldots.
\end{align}  
The negative binomial distribution is infinitely divisible in the first parameter. The cumulant generating function of $Z\sim \mathsf{NB}(r,p)$ is $r \ln \left( \frac{1-p}{1-pe^\theta} \right)$. 

We define the negative binomial channel as follows. 

\begin{definition}[negative binomial channel] \label{def:nb-channel}
Set $\kappa_{\mathsf{NB}}(\theta) = \ln \left( \frac{1/2}{1 - e^\theta /2} \right)$. Then, the discrete-time L\'evy channel generated by $\kappa_{\mathsf{NB}}(\theta)$ at SNR $\gamma$ is given by
\begin{equation}
Y_\gamma|X \sim \mathsf{NB}(\gamma, \frac{X}{1+X}), \quad X\geq 0. 
\end{equation}
\end{definition}

Indeed, we have $X = \kappa_{\mathsf{NB}}'(\theta) = \frac{\frac{1}{2}e^\theta}{1-\frac{1}{2}e^\theta}, \theta = \ln \left( \frac{2X}{1+X} \right)$. It follows from Lemma~\ref{lemma.outputcgf} that the cumulant generating function of the output is
\begin{align}
\ln \mathbb{E}_{Y_\gamma|X} e^{s Y_\gamma} & = \gamma \left( \ln \left( \frac{1/2}{1-e^{\theta +s}/2} \right) - \ln \left( \frac{1/2}{1-e^\theta /2} \right) \right) \\
& = \gamma \ln \left( \frac{1-\frac{X}{1+X}}{1- e^s \frac{X}{1+X}} \right),
\end{align}
which follows the distribution $\mathsf{NB}(\gamma, X/(1+X))$. 

Now we compute the L\'evy characteristics corresponding to $\kappa_{\mathsf{NB}}(\theta)$. We have
\begin{align}
\kappa_{\mathsf{NB}}(\theta) & = \ln\left(\frac{1}{2}\right) - \ln\left( 1-\frac{e^\theta}{2}\right) \\
& = \ln\left(\frac{1}{2}\right) - \sum_{i = 1}^\infty (-1)^{i+1} \frac{1}{i} \left( - \frac{e^\theta}{2} \right)^i \\
& = \ln\left(\frac{1}{2}\right) + \sum_{i = 1}^\infty e^{i\theta} \frac{1}{i 2^i} \\
& = \sum_{i = 1}^\infty  \left( e^{i\theta} -1\right) \frac{1}{i 2^i} \\
& = \int_{z\in \mathbb{N}_+} (e^{\theta z}-1) \nu(dz), 
\end{align}
where $\nu(k) = \frac{1}{k 2^k}, k \in \mathbb{N}_+$. It is clear that $\sigma = 0$.

Specializing Theorem~\ref{thm:levy-i-mmle} to the negative binomial channel leads to the following. 
\begin{theorem}
Let non-negative random variable $X$ satisfy $\E  X \ln \left| \frac{2X}{1+X} \right|<\infty$. Let $Y_\gamma$ denote the output of the negative binomial channel with input $X$ at SNR level $\gamma$. Then,
\begin{align}
\frac{\partial }{\partial \gamma} I(X; Y_\gamma) = \E[ \ell_{\mathsf{NB}} (X, \hat{X}^P_\gamma) ],
\end{align}
where the loss function $\ell_{\mathsf{NB}}$ is defined as,
\begin{align} \label{eq:nb-loss}
\ell_{\mathsf{NB}} (x, \hat{x}) = \sum_{z\in \mathbb{N}_+}  \ell_\Pscr(e^{z\ln \left ( 2x/(1+x)  \right ) }, \hat{x}_z) \frac{1}{z 2^z},
\end{align}
and the reconstruction $\hat{X}^P_\gamma = \{\hat{X}^P_{\gamma,z}: z\geq 1, z\in \mathbb{N}\} \in \mathcal{X}$ satisfies
\begin{align}\label{eqn.nbestimatorsdef}
\hat{X}^P_{\gamma,z} = \mathbb{E} \left[  \left( \frac{2X}{1+X}  \right)^z \Bigg|Y_\gamma\right],\quad \forall z\geq 1, z\in \mathbb{N}. 
\end{align}
Recall that $\ell_\Pscr$ is the Poisson loss function introduced in (\ref{eq: poisson loss}).
\end{theorem}

Analogous to our discussion so far, we now present the corresponding result for mismatched estimation.
\begin{theorem} \label{thm:nb-d-mle}  
Let non-negative random variable $X$ be the input to the negative binomial channel, and conditions in Theorem~\ref{thm:D-mle} are satisfied. Let $P_{Y_\gamma}$ and $Q_{Y_\gamma}$ denote the output laws when the input is distributed according to $P$ and $Q$ respectively. Then, we have,
\begin{align}
D(P_{Y_\gamma} || Q_{Y_\gamma}) = \int_0^\gamma \E_P[ \ell_{\mathsf{NB}} (X, \hat{X}^Q_\alpha) - \ell_{\mathsf{NB}} (X, \hat{X}^P_\alpha) ] \,d\alpha.
\end{align}
The loss function $\ell_{\mathsf{NB}}$ is defined in (\ref{eq:nb-loss}). The reconstructions $\hat{X}^P_\alpha$ and $\hat{X}^Q_\alpha$ are defined in~(\ref{eqn.nbestimatorsdef}). 
\end{theorem}

The so called negative binomial channel has appeared in the literature before, in the context of relations between information and estimation, cf. \cite{taborda2014information}. However, our approach is quite different from existing approaches, and in our point of view, much more natural. We will illustrate the key differences between our approach and existing approaches in negative binomial channels in the next section.



\section{Channels and loss functions: a discussion} \label{sec: bregman}

\subsection{Special case of deterministic inputs}

Applying Lemma~\ref{lemma.regular} to discrete-time L\'evy channels, we obtain a closed form representation of the relative entropy. 
\begin{corollary}\label{cor.levyexp}
Let $P^\gamma_{x}$ denote the output distribution of the discrete-time L\'evy channel generated by $\kappa(\theta)$ with input $x$ at SNR $\gamma$. Then, for two deterministic values $x_1,x_2$, 
\begin{align}
\frac{\partial}{\partial \gamma} D(P_{x_1}^\gamma \| P_{x_2}^\gamma) = \phi(x_1) - \phi(x_2) - \phi'(x_2)(x_1  -x_2).
\end{align}
\end{corollary}
\begin{IEEEproof}
Noting that the Fenchel--Legendre transform of $\gamma \kappa(\theta)$ is $\gamma \phi(x/\gamma)$, and the expectation of the outputs of the discrete-time L\'evy channels are $\gamma \kappa'(\theta_i) = \gamma x_i$, we have
\begin{align}
D(P_{x_1}^\gamma \| P_{x_2}^\gamma) & = \gamma \left( \phi \left( \frac{\gamma x_1}{\gamma} \right) - \phi\left( \frac{\gamma x_2}{\gamma} \right) - \frac{1}{\gamma} \phi'\left( \frac{\gamma x_2}{\gamma} \right) \left( \gamma x_1 - \gamma x_2 \right) \right)\\
& = \gamma \left( \phi(x_1) - \phi(x_2) - \phi'(x_2)(x_1  -x_2) \right). 
\end{align}
The claim follows by taking the derivative on both sides with respect to $\gamma$. 
\end{IEEEproof}

In retrospect, it is Corollary~\ref{cor.levyexp} that motivated us to define the discrete-time L\'evy channels, and extend nearly all aspects of information-estimation results in the literature about Gaussian and Poisson models to these channels. 

Interesting, Theorem~\ref{thm:D-mle} also contains the deterministic inputs as a special case. Specializing Theorem~\ref{thm:D-mle} to the case of $P = \delta_{x_1}, Q = \delta_{x_2}$ and comparing with Corollary~\ref{cor.levyexp}, we obtain:
\begin{theorem}\label{thm.lossbreg}
Let $P^\gamma_{x}$ denote the output distribution of the discrete-time L\'evy channel generated by $\kappa(\theta)$ with input $x$ at SNR $\gamma$. Then, for two deterministic values $x_1,x_2$, 
\begin{align}
\frac{\partial}{\partial \gamma} D(P_{x_1}^\gamma \| P_{x_2}^\gamma) & = \sigma^2 \ell_{\Gscr} (\phi'(x_1), \phi'(x_2)) + \int_{\R} \ell_{\Pscr} (e^{\phi'(x_1)z}, e^{\phi'(x_2)z})\, \nu(dz) \\
& = \phi(x_1) - \phi(x_2) - \phi'(x_2)(x_1  -x_2).
\end{align}
\end{theorem}

\begin{IEEEproof}
We give a direct proof below. Denoting $\theta_1 = \phi'(x_1), \theta_2 = \phi'(x_2)$, it follows from Definition~\ref{def.discrete-timelevy} that 
\begin{align}
\ln \frac{dP^\gamma_{x_1}}{dP^\gamma_{x_2}} & = \theta_1 Y_\gamma - \gamma \kappa(\theta_1) - \left( \theta_2 Y_\gamma - \gamma \kappa(\theta_2) \right). 
\end{align}
Taking expectation on both sides with respect to $P^\gamma_{x_1}$ and utilizing the fact that $\mathbb{E}_{P^\gamma_{x_1}} Y_\gamma = \gamma \kappa'(\theta_1)$, we have
\begin{align}
D(P^\gamma_{x_1} \| P^\gamma_{x_2}) & = \theta_1 \gamma \kappa'(\theta_1) - \gamma \kappa(\theta_1) - \theta_2 \gamma \kappa'(\theta_1) + \gamma \kappa(\theta_2) \\
& = \gamma \left( \kappa(\theta_2) - \kappa(\theta_1) - \kappa'(\theta_1)(\theta_2 - \theta_1) \right),
\end{align}
implying that
\begin{align}
\frac{\partial}{\partial \gamma} D(P^\gamma_{x_1} \| P^\gamma_{x_2}) & = \kappa(\theta_2) - \kappa(\theta_1) - \kappa'(\theta_1)(\theta_2 - \theta_1). 
\end{align}
Now, it suffices to show that
\begin{align}
\kappa(\theta_2) - \kappa(\theta_1) - \kappa'(\theta_1)(\theta_2 - \theta_1) & = \sigma^2 \ell_{\Gscr} (\phi'(x_1), \phi'(x_2)) + \int_{\R} \ell_{\Pscr} (e^{\phi'(x_1)z}, e^{\phi'(x_2)z})\, \nu(dz),
\end{align}
for $\theta_i = \phi'(x_i)$. 

Applying the representation~(\ref{eqn.cdfnewrepre}), since the Bregman divergence is a linear operator on the function $\kappa$ and maps affine functions to zero, we have
\begin{align}
\kappa(\theta_2) - \kappa(\theta_1) - \kappa'(\theta_1)(\theta_2 - \theta_1) & = \frac{1}{2}\sigma^2 (\theta_2 - \theta_1)^2 + \int_{\mathbb{R}} \left( e^{\theta_2 z} - e^{\theta_1 z} - z e^{\theta_1 z} (\theta_2 - \theta_1) \right) \nu(dz) \\
& = \sigma^2 \ell_{\Gscr}(\theta_1,\theta_2) + \int_{\mathbb{R}} \left( e^{\theta_1 z} \ln \left( \frac{e^{\theta_1 z}}{e^{\theta_2 z}} \right) - e^{\theta_1 z} + e^{\theta_2 z} \right) \nu(dz) \\
& = \sigma^2 \ell_{\Gscr}(\theta_1,\theta_2)  + \int_{\R} \ell_{\Pscr} (e^{\theta_1 z}, e^{ \theta_2 z})\, \nu(dz) \\
& = \sigma^2 \ell_{\Gscr} (\phi'(x_1), \phi'(x_2)) + \int_{\R} \ell_{\Pscr} (e^{\phi'(x_1)z}, e^{\phi'(x_2)z})\, \nu(dz). 
\end{align}
\end{IEEEproof}

Theorem~\ref{thm.lossbreg} shows that under deterministic inputs, the seemingly convoluted loss function for L\'evy channels collapses to a crisp closed form formula, which in turn is simply the Bregman divergence generated by the convex function $\phi(x)$.

Now, we specialize Theorem~\ref{thm.lossbreg} to the gamma and negative binomial channels for a simple representation of their respective loss functions, which were defined in Section \ref{sec: gamma}. 

For the gamma channel, we have $\kappa_{\Gamma}(\theta) = -\ln(1-\theta)$, whose Fenchel--Legendre transform is given by,
\begin{equation}
\phi_{\Gamma}(x) = x - 1 - \ln x, x>0.
\end{equation}
Thus, we have,
\begin{align}
 d_{\Gamma}(x_1,x_2)  = \frac{x_1}{x_2} - \ln \left( \frac{x_1}{x_2}\right) - 1.
\end{align}

\begin{figure}[htb]
\begin{minipage}[b]{1.0\linewidth}
  \centering
  \centerline{\epsfig{figure=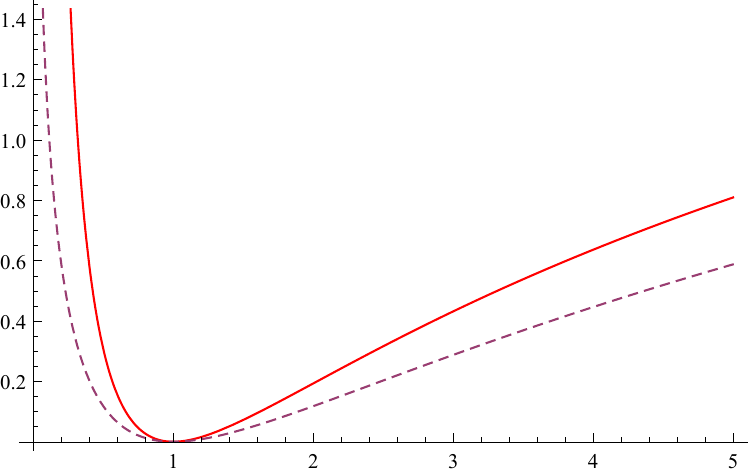,height=5.5cm}}
\end{minipage}
\centering
\caption{Visualization of $d_{\mathsf{NB}}(1,x)$ and $d_\Gamma(1,x)$. The solid curve is $d_\Gamma(1,x)$. The dashed curve is $d_{\mathsf{NB}}(1,x)$. }
\label{fig.nbandsato}
\end{figure}

The loss function $ d_{\Gamma}(x_1,x_2) $ is also called Itakura-Saito distance \cite{Banerjee--Merugu--Dhillon--Ghosh2005}, and has proved to play an important role in linear inverse problems as investigated by Csisz\'ar \cite{Csiszar1991}. To visualize this loss function, we fix $x_1 = 1$ and vary $x_2$ to obtain the solid curve in Figure~\ref{fig.nbandsato}. 

For negative binomial channels, we have $\kappa_{\mathsf{NB}}(\theta) =  \ln  \left( \frac{\frac{1}{2}}{1-\frac{1}{2}e^\theta} \right)$. The Fenchel--Legendre dual of $\kappa_{\mathsf{NB}}(\theta)$ is
\begin{equation}
\phi_{\mathsf{NB}}(x) = x \ln x - (1+x)\ln(1+x) + x \ln 2 + \ln 2,\quad  x\geq 0.
\end{equation}

Hence, 
\begin{align}
d_{\mathsf{NB}}(x_1,x_2) = x_1 \ln \left( \frac{x_1}{x_2} \right) + (1+x_1)\ln \left( \frac{1+x_2}{1+x_1}\right).
\end{align}

To visualize $d_{\mathsf{NB}}(x_1,x_2)$, we fix $x_1 = 1$ and vary $x_2$ to obtain the dashed curve in Figure~\ref{fig.nbandsato}. 

\subsection{Relations to other generalizations}

%
%

We briefly discuss the key differences between our approach and existing approaches to establishing information-estimation results in channels beyond Gaussian and Poisson. From now on we discuss the general natural exponential families introduced in Section~\ref{subsec.nef} and do not constrain ourselves to the discrete-time L\'evy channels. Targeting at several members of the natural exponential family, \cite{taborda2014information} took a view different from ours. They called an input amplification factor $a$ to be the parameter of interest. In other words, suppose $ax_1 = \kappa'(\theta_1), ax_2 = \kappa'(\theta_2), a> 0$, and consider deterministic inputs $ax_1,ax_2$, we have
\begin{equation}\label{eqn.apara}
D(P_1 \| P_2 ) = \phi(ax_1) - \phi(ax_2) - \phi'(ax_2)(ax_1 - ax_2),
\end{equation}
and the parameter $a$ is the parameter with respect to which \cite{taborda2014information} analyze the derivatives of mutual information and relative entropy.

Essentially, \cite{taborda2014information} worked on generalizing (\ref{eqn.apara}) to random inputs $X$.  However, it seems to the authors that even in the deterministic inputs case (\ref{eqn.apara}), the parameter $a$ may not display consistent properties for different channel models. Indeed, if we take derivatives on both sides of (\ref{eqn.apara}), we will obtain
\begin{equation}
a \frac{\partial}{\partial a} D(P_1 \| P_2 )  = d_{f}(ax_1,ax_2),
\end{equation}
where $f(x) = x\phi'(x), d_f(x,y) = f(x) - f(y) - f'(y)(x-y)$. However, $\phi(x)$ being a convex function does not imply $f(x) = x\phi'(x)$ is also a convex function, hence $d_f(x,y)$ may not be a Bregman divergence, and is not necessarily non-negative. Coincidentally, for Gaussian and Poisson models, the Bregman divergence generated by $x\phi'(x)$ and $\phi(x)$ are equal up to a multiplicative factor. Indeed, we have
\begin{align}
\phi(x) = \frac{1}{2}x^2, & \quad d_{\phi}(x,y) = \frac{1}{2}(x-y)^2, \\
x\phi'(x) = x^2, & \quad d_{x\phi'(x)}(x,y) = (x-y)^2,
\end{align}
for the Gaussian model, and
\begin{align}
\phi(x) = x\ln x - x + 1, & \quad d_{\phi}(x,y) = x \ln \left( \frac{x}{y} \right) - x+ y,\\
x\phi'(x) = x\ln x, & \quad d_{x\phi'(x)}(x,y) =  x \ln \left( \frac{x}{y} \right) - x+ y,
\end{align}
for the Poisson model. 

The Bregman divergences listed in \cite{taborda2014information} for binomial and negative binomial models are all Bregman divergences generated by $x\phi'(x)$, and it so happens that they are both strictly convex. However, if we consider the gamma distribution, we have
\begin{equation}
\phi_{\Gamma}(x) = x -1 -\ln(x), 
\end{equation}
which implies
\begin{equation}
f(x) = x \phi_{\Gamma}'(x) = x-1, \quad d_f(x,y) \equiv 0. 
\end{equation}

As a consequence of this parametrization, results in \cite{taborda2014information} do not take the forms of (\ref{eq: i-mmse}) and (\ref{eq: poisson i-mmle}). 

Further, we can show that even if we consider random inputs in the gamma distribution, if we follow the definition of parameter $a$ in \cite{taborda2014information}, we would obtain that the mutual information between input and output is \emph{invariant} with respect to the parameter $a$. In our definition of the gamma channel, we have
\begin{equation}
Y_\gamma|X \sim \Gamma(\gamma,X),
\end{equation}
however, if we follow \cite{taborda2014information}, then we have
\begin{equation}
Y_a|X \sim \Gamma(k,aX/k),
\end{equation}
where $k$ is some fixed positive constant. 

\begin{lemma}\label{lemma.gammaa}
Suppose $P,Q$ are two probability measures of non-negative random variable $X$. If we have $Y_a|X \sim \Gamma(k, aX/k), k>0$, then
\begin{align}
\frac{\partial}{\partial a}I(X; Y_a) & \equiv 0\\
\frac{\partial}{\partial a}D(P_{Y_a} \| Q_{Y_a}) & \equiv 0.
\end{align}
\end{lemma}

Lemma~\ref{lemma.gammaa} shows the parameterization in \cite{taborda2014information} may lead to some strange results that do not capture the infinite-divisibility of the gamma distribution. 

We hope to have convinced the reader that the parametrization in our framework is natural and captures the core properties of the Gaussian and Poisson distributions. In fact, we conjecture that discrete-time L\'evy channels are the largest family of channels for which one can establish information-estimation results paralleling all existing results in the Gaussian and Poisson observation models.

\section{Proofs} \label{sec: proofs}

\subsection{Proof of Theorem~\ref{thm:levy-i-mmle}}

Theorem~\ref{thm:levy-i-mmle} can be obtained via direct application of Theorem~\ref{thm:D-mle}. Indeed, mutual information $I(X;Y_\gamma)$ is expressible as
\begin{equation}
I(X; Y_\gamma) = \E_X D(P_{Y_\gamma|X} \| P_{Y_\gamma}),
\end{equation}
where $P_{Y_\gamma|X}$ is the marginal distribution of output of the L\'evy channel under point mass input $\delta_X$, and $P_{Y_\gamma}$ is the marginal distribution of $Y_\gamma$ under input $P_X$. 

It suffices to verify that the conditions of Theorem~\ref{thm:D-mle} are satisfied for $P$ corresponding to point mass $\delta_X$ and $Q$ corresponding to input distribution $P_X$. Indeed, it suffices to show that with probability one,
\begin{align}
\int_0^\gamma \mathbb{E}_{Y_\alpha|X} \left( \phi'(X) - \E[\phi'(X)|Y_\alpha] \right)^2 d\alpha & <\infty \\
\int_0^\gamma \int_{\mathbb{R}} \mathbb{E}_{Y_\alpha|X}   \left( e^{\phi'(X)z} \Bigg|\ln \left( \frac{e^{\phi'(X)z}}{\E_P[e^{\phi'(X)z}|Y_\alpha]} \right)\Bigg| \right) \nu(dz)d\alpha & <\infty. 
\end{align}

We have the following lemma. 
\begin{lemma}\label{lemma.verifycondition}
\begin{enumerate}
\item Suppose $\sigma\neq 0$ and $\mathbb{E}_P (\phi'(X))^2 <\infty$. Then with probability one for any $0<\gamma<\infty$, 
\begin{align}
\int_0^\gamma \mathbb{E}_{Y_\alpha|X} \left( \phi'(X) - \E[\phi'(X)|Y_\alpha] \right)^2 d\alpha & <\infty. 
\end{align}
\item Suppose $\int_{\R} \nu(dz)<\infty$ and $\mathbb{E}_P \int_{\R} |\phi'(X)z|e^{\phi'(X)z}\nu(dz)<\infty$. Then with probability one for any $0<\gamma<\infty$, 
\begin{align}
\int_0^\gamma \int_{\mathbb{R}} \mathbb{E}_{Y_\alpha|X}   \left( e^{\phi'(X)z} \Bigg|\ln \left( \frac{e^{\phi'(X)z}}{\E_P[e^{\phi'(X)z}|Y_\alpha]} \right)\Bigg| \right) \nu(dz)d\alpha & <\infty. 
\end{align}
\item Suppose $\int_{\R} \nu(dz) = \infty$, $\mathbb{E}_P \int_{\R} |\phi'(X)z|e^{\phi'(X)z}\nu(dz)<\infty$, and 
\begin{align}
\int_0^\gamma \int_{\mathbb{R}} \mathbb{E}   \left( e^{\phi'(X)z} \Bigg|\ln \E_P[e^{\phi'(X)z}|Y_\alpha]\Bigg| \right) \nu(dz)d\alpha & <\infty. 
\end{align} 
Then with probability one for any $0<\gamma<\infty$, 
\begin{align}
\int_0^\gamma \int_{\mathbb{R}} \mathbb{E}_{Y_\alpha|X}   \left( e^{\phi'(X)z} \Bigg|\ln \left( \frac{e^{\phi'(X)z}}{\E_P[e^{\phi'(X)z}|Y_\alpha]} \right)\Bigg| \right) \nu(dz)d\alpha & <\infty. 
\end{align}
\end{enumerate}
\end{lemma}
\begin{IEEEproof}
\begin{enumerate}
\item The desired claim follows from showing
\begin{align}
\int_0^\gamma \mathbb{E} (\phi'(X) - \mathbb{E}[\phi'(X)|Y_\alpha])^2 d\alpha <\infty,
\end{align}
which is implied by observing that $\mathbb{E} (\phi'(X) - \mathbb{E}[\phi'(X)|Y_\alpha])^2 \leq \mathbb{E} (\phi'(X))^2$.
\item It suffices to prove that
\begin{align}
\int_0^\gamma \int_{\mathbb{R}} \mathbb{E}  \left( e^{\phi'(X)z} \Bigg|\ln \E_P[e^{\phi'(X)z}|Y_\alpha]\Bigg| \right) \nu(dz)d\alpha & <\infty. 
\end{align}
We have
\begin{align}
\int_0^\gamma \int_{\mathbb{R}} \mathbb{E}  \left( e^{\phi'(X)z} \Bigg|\ln \E_P[e^{\phi'(X)z}|Y_\alpha]\Bigg| \right) \nu(dz)d\alpha & = \int_0^\gamma \int_{\mathbb{R}} \mathbb{E}  \left( \mathbb{E}_P[e^{\phi'(X)z}|Y_\alpha] \Bigg|\ln \E_P[e^{\phi'(X)z}|Y_\alpha]\Bigg| \right) \nu(dz)d\alpha \\
\end{align}
The claim follows from observing that 
\begin{enumerate}
\item For any $x\geq 0$, $x|\ln x| \leq -2x\ln x \mathbbm{1}_{0\leq x\leq 1} + x\ln x$;
\item For $x\in [0,1]$, $0\leq -2x\ln x \leq \frac{2}{e}$; 
\item $\int_{\R} \nu(dz) < \infty$;
\item The function $x\ln x$ is convex on $\mathbb{R}_+$; hence $\mathbb{E}[X] \ln (\mathbb{E}[X]) \leq \mathbb{E}[X \ln X]$; 
\item $\mathbb{E}_P \int_{\R} |\phi'(X)z|e^{\phi'(X)z}\nu(dz)<\infty$.
\end{enumerate}
\item It follows from the triangle inequality. 
\end{enumerate}
\end{IEEEproof}

Thus,
\begin{equation}
I(X; Y_\gamma) = \int_0^\gamma \E_P \ell_{\mathcal{L}}(X, \hat{X}_\alpha^P) d\alpha.
\end{equation}

The final claim follows from taking derivatives on both sides with respect to $\gamma$. 

\subsection{Proof of Theorem~\ref{thm:h-mmle}}

According to Theorem~\ref{thm:levy-i-mmle}, for any $\gamma$, 
\begin{equation}
I(X;Y_\gamma) =  \int_0^\gamma \E_P \ell_{\mathcal{L}}(X, \hat{X}_\alpha^P)d\alpha.
\end{equation}
Taking $\gamma \to \infty$ on both sides, we have
\begin{equation}
\lim_{\gamma \to \infty} I(X;Y_\gamma) =  \int_0^\infty \E_P \ell_{\mathcal{L}}(X, \hat{X}_\alpha^P)d\alpha,
\end{equation}
For discrete random variables $X$, 
\begin{equation}
\lim_{\gamma \to \infty} I(X;Y_\gamma) = H(X).
\end{equation}

\subsection{Proof of Theorem~\ref{thm:D-mle}}

%

Applying the coupling in Lemma~\ref{lemma.rdderivative}, we define the non-negative martingale process with respect to filtration $\mathcal{F}_t = \Fscr_t^Y \vee \sigma\{X\}$:
\begin{equation}
L_t \doteq \frac{dP^{[0,t]}_\theta}{dP^{[0,t]}_0}(Y_s, 0\leq s\leq t) = e^{\theta Y_t - t \kappa(\theta)},
\end{equation}
where $\{Y_s:0\leq s\leq t\}$ follows distribution $P^{[0,t]}_0$. The random variable $\theta = \phi'(X)$ is a function of the input $X$ (see Definition~\ref{def.discrete-timelevy}). Note that $A \vee B$ denotes the smallest $\sigma$-algebra that contains both $\sigma$-algebras $A$ and $B$, and $\mathcal{F}_t^Y$ denotes the filtration $\sigma\{Y_s:0\leq s\leq t\}$.

Applying~(\ref{eqn.cdfnewrepre}) in Lemma~\ref{lemma.lemmakhintchine},  Lemma~\ref{lemma.levyitodecomposition} and plugging those in the expression of $L_t$, we have
\begin{align}
L_t & = e^{\theta \left(at + \sigma W_t + \int_0^t \int_{|z|< 1} z (\mu(ds,dz) - \nu(dz)ds) + \int_0^t \int_{|z| \geq 1} z \mu(ds,dz) \right)- t \kappa(\theta)} \\
& = e^{\sigma \theta W_t - \frac{1}{2}\sigma^2 \theta^2 t + \int_0^t \int_{\mathbb{R}} \left( \theta z \mu(ds,dz) - (e^{\theta z}-1) \nu(dz)ds \right)}. 
\end{align} 

We have the following It$\hat{\mathrm{o}}$'s formula for general semimartingales, and we refer the readers to~\cite{medvegyev2007stochastic} for the general theory of semimartingales. 
\begin{lemma}\cite[Thm. 6.46]{medvegyev2007stochastic}\label{lemma.itosemimg}
If $\{Z(t):t\geq 0\}$ is a semimartingale and $f(x) \in C^2(\mathbb{R})$, then
\begin{align}
f(Z(t)) - f(Z(0)) & = \int_0^t f'(Z_{-})dZ + \frac{1}{2} \int_0^t f''(Z_{-}) d[Z]^c + \sum_{0<s\leq t} \left( f(Z(s)) -f(Z(s-)) - f'(Z(s-))\Delta Z(s) \right),
\end{align}
where the process $[Z]^c_t$ is the quadratic variation process of the continuous part of the semimartingale $Z(t)$, $\Delta Z(s) = Z(s) - Z(s-)$, and $Z(s-) = \lim_{u\to s-} Z(u)$. 
\end{lemma}
Setting
\begin{align}
Z(t) & = \sigma \theta W_t - \frac{1}{2}\sigma^2 \theta^2 t + D(t)\\
D(t) & = \int_0^t \int_{\mathbb{R}} \left( \theta z \mu(ds,dz) - (e^{\theta z}-1) \nu(dz)ds \right) \\
f(z) & = e^{z},
\end{align}
we know
\begin{align}
[Z]_t^c & = \sigma^2 \theta^2 t. 
\end{align}

Applying Lemma~\ref{lemma.itosemimg}, we get the following representation of the martingale $L_t$:
\begin{align}
L_t & = 1 + \int_0^t L_{s-} dZ(t) + \frac{1}{2}  \sigma^2 \theta^2 \int_0^t L_{s-} ds + \sum_{0<s\leq t} f(Z(s-)) \left( \frac{f(Z(s))}{f(Z(s-))} - 1 - \Delta Z(s) \right) \\
& = 1 + \int_0^t \sigma \theta L_{s-} dW_s + \int_0^t L_{s-} dD(s) + \sum_{0<s\leq t} L_{s-} \left( e^{\Delta Z(s)} - 1 - \Delta Z(s) \right) \\
& = 1 + \int_0^t \sigma \theta L_{s-} dW_s + \int_0^t L_{s-} dD(s) + \sum_{0<s\leq t} L_{s-} \left( e^{\theta z } - 1 - \theta z  \right) \mu(ds,dz) \\
& = 1 + \int_0^t L_{s-} dM_t,\label{eqn.sde}
\end{align} 
where 
\begin{align}
M_t & = \sigma \theta W_t + \int_0^t \int_{\mathbb{R}} \left( \theta z \mu(ds,dz) - (e^{\theta z}-1)\nu(dz)ds  + e^{\theta z}\mu(ds,dz) - (1+\theta z)\mu(ds,dz) \right) \\
& = \sigma \theta W_t + \int_0^t \int_{\mathbb{R}} (e^{\theta z}-1) \left( \mu(ds,dz) - \nu(dz)ds \right). 
\end{align}

The equation~(\ref{eqn.sde}) is called the Dol\'eans--Dade equation~\cite[Def. 6.53]{medvegyev2007stochastic}, whose solution is unique and is given by
\begin{equation}\label{eqn.solsde1}
L_t = e^{\varphi_t},
\end{equation}
where 
\begin{equation}\label{eqn.solsde2}
\varphi_t = M_t - \frac{1}{2}[M]_t^c + \sum_{s\leq t} \left( \ln(1+\Delta M_s) - \Delta M_s \right).
\end{equation}

Plugging $M_t$ into the general solution, one can also verify that $L_t$ is indeed the unique solution.

In order to characterize the relative entropy, we need to compute the marginal distribution of $Y_t$. Let $P_{Y_t}$ denote the probability measure of $\{Y_s:0\leq s\leq t\}$ with respect to the filtration $\mathcal{F}_t^Y$ when the input $X$ has distribution $P$, $Q_{Y_t}$ denote that when the input has distribution $Q$, and $R_{Y_t}$ denote that when there the input corresponds to $\theta = 0$. 

We have the following lemma. 
\begin{lemma}\label{lemma.conditionalexp}
Given probability space $(\Omega, \mathcal{F}, \mathbb{P})$, consider another probability measure $Q$ such that $Q \ll P$. Denote the Radon--Nikodym derivative as $Z = \frac{d \mathbb{Q}}{ d \mathbb{P}}$. Then,
\begin{enumerate}
\item For sub-$\sigma$-algebra $\mathcal{G} \subset \mathcal{F}$, we have
\begin{align}
\frac{d \mathbb{Q}_{|\mathcal{G}}}{d\mathbb{P}_{|\mathcal{G}}} & = \mathbb{E}_P[Z|\mathcal{G}].
\end{align}
\item For any $\mathcal{F}$-measurable random variable $X$, we have
\begin{align}
 \mathbb{E}_P[ZX|\mathcal{G}] & =\mathbb{E}_P[Z|\mathcal{G}] \mathbb{E}_Q[X|\mathcal{G}]. 
\end{align}
\end{enumerate}
\end{lemma}

It follows from Lemma~\ref{lemma.conditionalexp} that
\begin{equation}
\frac{dP_{Y_t}}{dR_{Y_t}} = \E_{R} \left[ L_t | \Fscr_t^Y \right],\quad R_{Y_t}-a.s.
\end{equation}

Denote $\frac{dP_{Y_t}}{dR_{Y_t}} $ by $\bar{L}_t^P$. We have
\begin{equation}
L_t = 1 + \int_0^t \sigma L_{s-} \theta dW_s + \int_0^t L_{s-} \left( \int_{\mathbb{R}} (e^{\theta z} - 1)(\mu(ds,dz) - \nu(dz)ds) \right).  
\end{equation}

Taking conditional expectations with respect to $\Fscr_t^Y$ on both sides under $R_{Y_t}$, noting that $W_t$ and $\mu(dt,sz)$ are measurable with respect to $\mathcal{F}_t^Y$ and applying the Fubini-type theorem for conditional expectations in~\cite[Thm. 2]{Kailath--Segall--Zakai1978}, we have
\begin{equation}
\bar{L}_t^P = 1+ \int_0^t \sigma \E_R[L_{s-}\theta|\Fscr_s^Y] dW_s+ \int_0^t  \left( \int_{\mathbb{R}} (\E_R[L_{s-}(e^{\theta z}-1)|\Fscr_s^Y])(\mu(ds,dz) - \nu(dz)ds) \right).  
\end{equation}

It follows from Lemma~\ref{lemma.conditionalexp} that
\begin{align}
 \E_R[L_{s-}\theta|\Fscr_s^Y] & = \bar{L}^P_{s-}\E_P[\theta|\Fscr_s^Y]\\
 \E_R[L_{s-}(e^{\theta z}-1)|\Fscr_s^Y]) & = \bar{L}^P_{s-} (\E_P[e^{\theta z}|\Fscr_s^Y] - 1),
\end{align}
which implies
\begin{equation}
\bar{L}_t^P = 1+ \int_0^t \sigma \bar{L}^P_{s-} \E_P[\theta|\Fscr_s^Y] dW_s+ \int_0^t \bar{L}^P_{s-} \left( \int_{\mathbb{R}} (\E_P[e^{\theta z}|\Fscr_s^Y] - 1)(\mu(ds,dz) - \nu(dz)ds) \right).  
\end{equation}

Solving this stochastic differential equation using the general solutions given by~(\ref{eqn.solsde1}) and (\ref{eqn.solsde2}), we have
\begin{equation}
\bar{L}_t^P = e^{\rho_t^P},
\end{equation}
where
\begin{equation}
\rho_t^P = \sigma \int_0^t \E_P[\theta|\Fscr_s^Y] dW_s - \frac{1}{2}\sigma^2 \int_0^t (\E_P[\theta|\Fscr_s^Y])^2 ds + \int_0^t \int_{\mathbb{R}} \left( \ln \left(\E_P[e^{\theta z}|\Fscr_s^Y] \right) \mu(ds,dz) - (\E_P[e^{\theta z}|\Fscr_s^Y]-1)\nu(dz)ds \right).
\end{equation}

We have similar expressions for $\bar{L}_t^Q = \frac{dQ_{Y_t}}{dR_{Y_t}}$. In order to calculate the relative entropy, we have
\begin{align}
\ln \frac{dP_{Y_t}}{dQ_{Y_t}} & = \ln \left( \frac{dP_{Y_t}}{dR_{Y_t}} \frac{dR_{Y_t}}{dQ_{Y_t}} \right) \\
& = \ln \frac{dP_{Y_t}}{dR_{Y_t}} - \ln \frac{dQ_{Y_t}}{dR_{Y_t}} \\
& = \rho_t^P - \rho_t^Q \\
& = \sigma \int_0^t \left( \E_P[\theta|\Fscr_s^Y] - \E_Q[\theta|\Fscr_s^Y] \right)dW_s - \frac{1}{2}\sigma^2 \int_0^t \left( \E_P[\theta|\Fscr_s^Y]^2 - \E_Q[\theta|\Fscr_s^Y]^2 \right)ds  \\
& \quad + \int_0^t \int_{\mathbb{R}} \left( \ln \left( \frac{\E_P[e^{\theta z}|\Fscr_s^Y]}{\E_Q[e^{\theta z}|\Fscr_s^Y]} \right) \mu(ds,dz) - (\E_P[e^{\theta z}|\Fscr_s^Y] - \E_Q[e^{\theta z}|\Fscr_s^Y])\nu(dz)ds \right)
\end{align}

It follows from~\cite[Chap. 4, Sec. 6, Thm. 5]{Liptser--Shiryaev1989} that under probability measure $P_{Y_t}$, the process 
\begin{equation}
\tilde{W}_t = W_t - \int_0^t \E_P[\theta|\Fscr_s^Y] ds
\end{equation}
is a Brownian motion. Also, under $P_{Y_t}$, the compensator of $\mu(ds,dz)$ is no longer $\nu(dz)ds$, but $\E_P[e^{\theta z}|\Fscr_s^Y] \nu(dz)ds$. 

Bearing these in mind, we represent $\ln \frac{dP_{Y_t}}{dQ_{Y_t}}$ as
\begin{align}
\ln \frac{dP_{Y_t}}{dQ_{Y_t}} & = \sigma \int_0^t \left( \E_P[\theta|\Fscr_s^Y] - \E_Q[\theta|\Fscr_s^Y] \right)d\tilde{W}_s + \frac{1}{2}\sigma^2\int_0^t \left( \E_P[\theta|\Fscr_s^Y] - \E_Q[\theta|\Fscr_s^Y]\right)^2 ds \\
& \quad + \int_0^t \int_{\mathbb{R}} \ln \left( \frac{\E_P[e^{\theta z}|\Fscr_s^Y]}{\E_Q[e^{\theta z}|\Fscr_s^Y]} \right) \left( \mu(ds,dz) - \E_P[e^{\theta z}|\Fscr_s^Y] \nu(dz)ds \right) \\
& \quad + \int_0^t \int_{\mathbb{R}} \left(\E_P[e^{\theta z}|\Fscr_s^Y] \ln \left( \frac{\E_P[e^{\theta z}|\Fscr_s^Y]}{\E_Q[e^{\theta z}|\Fscr_s^Y]} \right) - \E_P[e^{\theta z}|\Fscr_s^Y] + \E_Q[e^{\theta z}|\Fscr_s^Y]  \right)\nu(dz)ds
\end{align}
 
Now we take expectations with respect to $P_{Y_t}$ on both sides. If the integrand of the stochastic integral with respect to $\tilde{W}_s$ is square integrable, i.e.
\begin{equation}
\E_P \int_0^t \left( \E_P[\theta|\Fscr_s^Y] - \E_Q[\theta|\Fscr_s^Y] \right)^2 ds<\infty,
\end{equation}
then
\begin{equation}
\sigma \int_0^t \left( \E_P[\theta|\Fscr_s^Y] - \E_Q[\theta|\Fscr_s^Y] \right)d\tilde{W}_s
\end{equation}
is a martingale, hence has mean zero. 

Meanwhile, according to \cite[Thm 18.7]{Liptser--Shiryaev2001}, if 
\begin{equation}
\int_0^t \int_{\mathbb{R}} \E_P \left( \E_P[e^{\theta z}|\Fscr_s^Y] \Bigg|\ln \left( \frac{\E_P[e^{\theta z}|\Fscr_s^Y]}{\E_Q[e^{\theta z}|\Fscr_s^Y]} \right)\Bigg| \right) \nu(dz)ds <\infty,
\end{equation}
then 
\begin{equation}
\int_0^t \int_{\mathbb{R}} \ln \left( \frac{\E_P[e^{\theta z}|\Fscr_s^Y]}{\E_Q[e^{\theta z}|\Fscr_s^Y]} \right) \left( \mu(ds,dz) - \E_P[e^{\theta z}|\Fscr_s^Y] \nu(dz)ds \right)
\end{equation}
is a zero mean martingale. 

Both conditions are guaranteed by the assumptions. Hence, 
\begin{align}
D(P_{Y_t} \| Q_{Y_t}) & = \frac{1}{2}\sigma^2\int_0^t \E_P \left( \E_P[\theta|\Fscr_s^Y] - \E_Q[\theta|\Fscr_s^Y]\right)^2 ds\\
& \quad +\int_0^t \int_{\mathbb{R}} \E_P \left(\E_P[e^{\theta z}|\Fscr_s^Y] \ln \left( \frac{\E_P[e^{\theta z}|\Fscr_s^Y]}{\E_Q[e^{\theta z}|\Fscr_s^Y]} \right) - \E_P[e^{\theta z}|\Fscr_s^Y] + \E_Q[e^{\theta z}|\Fscr_s^Y]  \right)\nu(dz)ds.\\
& = \E_P \int_0^t \ell_{\mathcal{L}}(X, \hat{X}_s^Q) - \ell_{\mathcal{L}}(X, \hat{X}_s^P) ds,
\end{align}
where in the last step we have used the following facts:
\begin{align}
\E_P (X - \E_Q[X|Y])^2 - \E_P(X - \E_P[X|Y])^2 & = \E_P \left( \E_P[X|Y] - \E_Q[X|Y] \right)^2,
\end{align}
\begin{align}
\E_P \ell_{\Pscr}(X, \E_Q[X|Y]) - \E_P \ell_{\Pscr}(X, \E_P[X|Y]) & =  \E_P \ell_{\Pscr}(\E_P[X|Y],\E_Q[X|Y]).
\end{align}

\subsection{Proof of Theorem~\ref{thm.levyrelativerepre}}

According to Theorem~\ref{thm:D-mle}, we know for all $\gamma<\infty$, 
\begin{equation}\label{eqn.monotogetorigin}
D(P_{Y_\gamma} || Q_{Y_\gamma}) = \int_0^\gamma \E_{P}[ \ell_{\Lscr} (X, \hat{X}^Q_\alpha)  - \ell_{\Lscr} (X, \hat{X}^P_\alpha)] \, d\alpha.
\end{equation}

We will first show that 
\begin{equation}
\lim_{\gamma \to \infty} D(P_{Y_\gamma} || Q_{Y_\gamma}) = D(P \| Q). 
\end{equation}

Note that
\begin{align}
D(P_{Y_\gamma} || Q_{Y_\gamma}) & \leq D(P_{X,Y_\gamma} \| Q_{X,Y_\gamma}) \\
& = D(P \| Q) + D(P_{Y_\gamma|X} \| Q_{Y_\gamma|X} | P) \\
& = D(P \| Q), 
\end{align}
where the last equality is due to the fact that $P_X$-a.s. $P_{Y_\gamma|X} = Q_{Y_\gamma|X}$ provided that $P \ll Q$. The monotonicity of $D(P_{Y_\gamma} || Q_{Y_\gamma})$ (which follows from~(\ref{eqn.monotogetorigin})) implies that the limit exists when $\gamma \to \infty$, and
\begin{equation}
\lim_{\gamma \to \infty }D(P_{Y_\gamma} || Q_{Y_\gamma})  \leq D(P \| Q). 
\end{equation}

On the other hand, by the Law of Large Numbers, $Y_\gamma/\gamma$ converges weakly to $X$ when $\gamma \to \infty$. Since relative entropy is lower semi-continuous under weak convergence, we have
\begin{align}
\liminf_{\gamma \to \infty} D(P_{Y_\gamma} || Q_{Y_\gamma}) & = \liminf_{\gamma \to \infty} D(P_{Y_\gamma /\gamma} \| Q_{Y_\gamma /\gamma}) \\
& \geq D(P \| Q).
\end{align}

We conclude that $\lim_{\gamma \to \infty} D(P_{Y_\gamma} || Q_{Y_\gamma}) = D(P \| Q)$.

Since
\begin{equation}
\lim_{\gamma \to \infty} \int_0^\gamma \E_{P}[ \ell_{\Lscr} (X, \hat{X}^Q_\alpha)  - \ell_{\Lscr} (X, \hat{X}^P_\alpha)] \, d\alpha = \int_0^\infty \E_P [ \ell_{\Lscr} (X, \hat{X}^Q_\alpha)  - \ell_{\Lscr} (X, \hat{X}^P_\alpha)] \, d\alpha,
\end{equation}
we have
\begin{equation}
D(P \| Q) =  \int_0^\infty \E_{P}[ \ell_{\Lscr} (X, \hat{X}^Q_\alpha)  - \ell_{\Lscr} (X, \hat{X}^P_\alpha)] \, d\alpha.
\end{equation}

%
%
%

\subsection{Proof of Lemma~\ref{lemma.gammaa}}

We first prove that, in the gamma distribution indexed by $\Gamma(k, aX/k)$, the following relationship holds:
\begin{equation}\label{eqn.dery}
a \frac{\partial p(y)}{\partial a} = -\frac{d (y p(y))}{d y}.
\end{equation}

Indeed, we have 
\begin{equation}
p(y) = \int_{\mathcal{X}} \frac{k^k y^{k-1} e^{-\frac{ky}{ax}}}{(ax)^k \Gamma(k)}dP_X,
\end{equation}
and
\begin{align}
a \frac{\partial p(y)}{\partial a}& = \int_{\mathcal{X}} \frac{a k^k y^{k-1}}{\Gamma(k)} \left(  \frac{e^{-\frac{ky}{ax}} \frac{ky}{xa^2}(ax)^k - k (ax)^{k-1} x e^{-\frac{ky}{ax}}}{(ax)^{2k}} \right )dP_X \\
& = \int_{\mathcal{X}} \frac{k^k y^{k-1} e^{-\frac{ky}{ax}}}{\Gamma(k)(ax)^k} \left( \frac{ky}{ax}-k\right) dP_X,
\end{align}
as well as
\begin{align}
-\frac{d (yp(y))}{dy} & = - \frac{d}{dy} \int_{\mathcal{X}} \frac{k^k y^k e^{-\frac{ky}{ax}}}{(ax)^k \Gamma(k)} dP_X \\
& = - \int_{\mathcal{X}} \frac{k^k e^{-\frac{ky}{ax}}}{(ax)^k \Gamma(k)} \left( k y^{k-1} - \frac{k y^k}{ax} \right)dP_X \\
& = \int_{\mathcal{X}} \frac{k^k y^{k-1} e^{-\frac{ky}{ax}}}{\Gamma(k)(ax)^k} \left( \frac{ky}{ax}-k\right) dP_X \\
& = a \frac{\partial p(y)}{\partial a}.
\end{align}

Based on (\ref{eqn.dery}), we have the following steps:

\begin{align}
a \frac{\partial D(P_Y \| Q_Y)}{\partial a} & = - \int \frac{d(y p(y))}{dy} \ln \frac{p(y)}{q(y)}dy + \int \frac{p(y)}{q(y)} d(yq(y)) \\
& = \int \ln q(y) d(y p(y)) - \int \ln p(y) d(y p(y)) +  \int \frac{p(y)}{q(y)} d(yq(y)) \\
& = - \int \frac{d q(y)}{q(y)} y p(y) + \int \frac{dp(y)}{p(y)} y p(y) + \int \frac{ p(y)}{q(y)} (y dq(y) + q(y)dy) \\
& = - \int y \frac{p(y)}{q(y)}dq(y) + \int y dp(y) + \int y \frac{p(y)}{q(y)}dq(y) + \int p(y)dy \\
& = \int y dp(y) + 1 \\
& = \int \left[ - \frac{\partial p(y)}{\partial a}  \right]dy - \int p(y)dy + 1 \\
& = \frac{\partial }{\partial a} \int p(y)dy \\
& = 0.
\end{align}

The result for mutual information follows from expressing $I(X;Y_a)$ via
\begin{equation}
I(X;Y_a) = D(P_{Y_a|X} \| P_{Y_a}|P_X). 
\end{equation}

\subsection{Proof of Lemma~\ref{lemma.conditionalexp}}

The first claim follows from showing that for any $B\in \mathcal{G}$, 
\begin{align}
\mathbb{Q}(B) = \mathbb{E}_P[Z \mathbbm{1}_B] = \mathbb{E}_P[\mathbb{E}_P[Z\mathbbm{1}_B|\mathcal{G}]] = \mathbb{E}_P[ \mathbb{E}_P[Z|\mathcal{G}] \mathbbm{1}_{B}] = \int_B \mathbb{E}_P[Z|\mathcal{G}] d\mathbb{P}. 
\end{align}
Regarding the second claim, denote $Y = \mathbb{E}_P[Z|\mathcal{G}]$. Then, for any $B\in \mathcal{G}$, we have
\begin{align}
\int_B \mathbb{E}_P[ZX|\mathcal{G}] d\mathbb{Q} & = \int_B  Y \mathbb{E}_P[ZX|\mathcal{G}] d\mathbb{P} \\
& = \mathbb{E}_P [\mathbb{E}_P[Y \mathbbm{1}_B ZX |\mathcal{G}]] \\
& = \mathbb{E}_P[Y \mathbbm{1}_B ZX] \\
& = \mathbb{E}_Q[Y \mathbbm{1}_B X] \\
& = \mathbb{E}_Q [ \mathbb{E}_Q[Y \mathbbm{1}_B X|\mathcal{G}] ] \\
& = \mathbb{E}_Q [ Y \mathbbm{1}_B  \mathbb{E}_Q[X|\mathcal{G}] ] \\
& = \int_B Y \mathbb{E}_Q[X|\mathcal{G}] d\mathbb{Q}. 
\end{align}
Since $B\in \mathcal{G}$ is arbitrary, we have
\begin{align}
\mathbb{E}_P[ZX|\mathcal{G}]  & = Y  \mathbb{E}_Q[X|\mathcal{G}]. 
\end{align}

\section{Conclusions} \label{sec: conclusions}
We have introduced the family of discrete-time L\'evy channels, where the output conditioned on the input is a random variable having an infinitely divisible law. We establish new and general relations between fundamental information measures and optimal estimation loss for this class of channels, under natural and explicitly identified loss functions. We conjecture that the discrete-time L\'evy channels are the largest family of channels admitting information-estimation relations that fully parallel the known results for the Gaussian and Poisson models. It would be an interesting challenge to prove our main results without using the tools from continuous time stochastic process theory. 

\section*{Acknowledgment}

We are grateful to the associate editor and the anonymous reviewers for various constructive suggestions that helped significantly improve the presentation of the paper.


\bibliographystyle{IEEEtran}
\bibliography{di_inf_est}

\end{document}